\documentclass{article}
\usepackage{epsfig}
\usepackage{verbatim}
\usepackage{natbib}
\newcommand{\vs}{\textit{vs}.\ }

\newcommand{\Fig}[1]{Figure \ref{f:#1}}

\begin{document}
\begin{center}
{\LARGE Fragment properties at the catastrophic disruption threshold:\\}
{\LARGE The effect of the parent body's internal structure\\}
\vspace {5 cm}
{\large }
{by\\}{
\vspace {.5 cm}
MARTIN JUTZI$^{1}$, PATRICK MICHEL$^{2}$, WILLY BENZ$^{1}$, DEREK C. RICHARDSON$^{3}$ \\}
\vspace {1. cm}
{$^1$ Physikalisches Institut, University of Bern, Sidlerstrasse 5, CH-3012
Bern, Switzerland \\}
{$^2$ University of Nice-Sophia Antipolis,  UMR 6202 Cassiop\'ee/CNRS, Observatoire de la C\^ote d'Azur, 
B.P. 4229, 06304 Nice cedex 4,
France\\}
{$^3$ Department of Astronomy, University of Maryland, College Park, MD 20742-2421, USA\\}
\vspace{.5 cm}
TEL: (+41) 31 631 4429 \\
FAX: (+41) 31 631 4405 \\
E-MAIL: jutzi@space.unibe.ch\\
\end{center}
\vspace {5 cm}
\textbf{Length:\\} 18 manuscript pages\\5 Tables\\19 figures\\%

\newpage
\noindent\textbf{Running Title:\\} Fragment properties at the catastrophic disruption threshold\\\\
\textbf{Corresponding author:\\}
Martin Jutzi\\
Physikalisches Institut \\
University of Bern \\
Sidlerstrasse 5 \\
CH-3012 Bern \\
Switzerland\\
TEL: (+41) 31 631 4429 \\
FAX: (+41) 31 631 4405\\
E-MAIL: jutzi@space.unibe.ch \\

\newpage
\centerline{ABSTRACT}
\bigskip

Numerical simulations of asteroid break-ups, including both the fragmentation of
the parent body and the gravitational interactions between the fragments, have
allowed us to reproduce successfully the main properties of asteroid families
formed in different regimes of impact energy, starting from a non-porous parent body. In this paper, 
using the same approach, we concentrate on a single regime of impact energy, the
so-called catastrophic threshold usually designated by $Q^*_D$, which results
in the escape of half of the target's mass. Thanks to our recent implementation of 
a model of fragmentation of porous materials, we can characterize $Q^*_D$ 
for both porous and non-porous targets with a wide range of diameters. 
We can then analyze the potential influence of porosity on the value of $Q^*_D$, and by computing 
the gravitational phase of the collision in the gravity regime, we can characterize the collisional outcome in terms of the fragment size and ejection speed distributions, which are
the main outcome properties used by collisional models to study the evolutions
of the different populations of small bodies. We also check the dependency of $Q^*_D$ on the 
impact speed of the projectile.

In the strength regime, which corresponds to target sizes below a few hundreds of meters, we find that 
porous targets are more difficult to disrupt than non-porous ones. In the gravity regime, the outcome 
is controlled purely by gravity and porosity in the case of porous targets. In the case of non-porous targets, the outcome also depends on strength. Indeed, decreasing the strength of non-porous targets make them easier to disrupt in this regime, while increasing  
the strength of porous targets has much less influence on the value of $Q^*_D$. Therefore, one cannot say that non-porous 
targets are systematically easier or more difficult to disrupt than porous ones, as the outcome highly depends on the 
assumed strength values. In the gravity regime, we also confirm that the process of gravitational reaccumulation 
is at the origin of the largest remnant's mass in both cases. We then propose some power-law relationships between 
$Q^*_D$ and both target's size and impact speed that can be used in collisional evolution models. The resulting fragment size distributions can 
also be reasonably fitted by a power-law whose exponent ranges 
between $-2.2$ and $-2.7$ for all target diameters in both cases and independently on the impact velocity (at least 
in the small range investigated between 3 and 5 km$/$s). Then, although ejection velocities in the gravity regime tend to be higher 
from porous targets, they remain on the same order as the ones from non-porous targets.   

\newpage

\section{Introduction}

The collisional evolution of small body populations, such as the main belt
asteroids, is generally studied by using numerical models that compute the
evolution of the size and velocity distributions of objects as a result of
both collisional and dynamical processes. The asteroid disruption and
fragmentation algorithms used in such models contain significant
uncertainties. In particular, the scaling parameter is commonly
defined as the critical specific impact energy $Q^*_D$, which results
in the escape of half of the target's mass (hence the label $D$ for dispersal), called also the catastrophic
impact energy threshold. Thus, $Q^*_D$ is a critical function needed in all
codes that include fragmentation between rocky bodies. In this paper, by
numerically simulating the collisional process between small bodies, we look
for the value of this catastrophic threshold as a function of the  
target's diameter, internal structure, as well as impact speed, and we characterize the
outcome in terms of fragment size and ejection velocity
distributions. Ideally, a wide range of outcomes should be determined
depending on the collisional energy (i.e. not limited to the catastrophic
threshold), but this requires very massive computations that will be 
undertaken in further studies.

The specific impact energy is defined as $Q=0.5 m_p v^2_p/M_t$, where $m_p$, $v_p$ and
$M_t$ are the mass and speed of the projectile and the target's mass,
respectively. The catastrophic disruption threshold $Q^*_D$ is defined as the
specific impact energy leading to a largest fragment containing $50 \%$ of the original
target's mass. Up to now, some {\it scaling laws} have been used in
collisional models to define the outcome properties of a collision. These
scaling laws have been developed in order to extrapolate the results of
laboratory experiments on centimeter-size targets to asteroid scales. They
generally employ non-dimensional ratios involving projectile's size, impact
speed and target's strength, and they have led to the characterization of a
relationship between $Q^*_D$ and the target's diameter $D$ over a wide range
of values (from centimeters to several hundreds kilometers). Unfortunately, the
relations derived from these scaling laws assume a uniformity of process,
structural continuity and other idealizations that put their applicability into
question. As a result, depending on the assumptions made, the relationships
obtained between $Q^*_D$ and $D$ can differ over several orders of
magnitude. Nevertheless, despite these discrepancies, some systematic
trends arise. In particular, impacts on small objects take place in
the so-called ``strength-scaling'' regime, where the fragmentation of
an object is essentially governed by its tensile strength, whereas
impacts on large asteroids take place in the ``gravity-scaling''
regime, where the gravity is generally assumed to control the outcome.
Benz  and Asphaug (1999) found that the transition between the two regimes may 
occur in the range of target's diameters between $100$ and $200$ meters.
Values of $Q^*_D$ have been
estimated using both laboratory and numerical hydrocode experiments (see
recent reviews on these topics by Holsapple et al. 2002, and Asphaug et
al. 2002).

The first {\it self consistent} study aimed at characterizing the catastrophic
disruption threshold was performed by Benz and Asphaug (1999), who used a
smoothed particle hydrodynamics (SPH) code to simulate the break-up of
basalt and icy bodies from centimeters-scale to hundreds kilometers in
diameter. Their simulations included in a simplified way the combined effects of material
strength and self-gravitation, which allowed covering both the 
strength and the gravity regimes. However, in the latter
regime, their study was still limited by the fact that
the gravitational phase was not explicitly simulated; rather, an iterative
procedure based on energy balance was used to identify the largest
fragment formed by reaccumulation of smaller ones (see Benz and Asphaug 1999,
Section 3.3 for details). Nevertheless, the mass of large fragments can be determined quite accurately using this method.

Recently, Leinhardt and Stewart (2008) started to investigate the dependency of $Q^*_D$ 
on the strength of the body using the hydrocode CTH (McGlaun et al. 1990) to compute the fragmentation phase and the $N$-body 
code {\it pkdgrav} to compute the subsequent gravitational evolution of the fragments. They found that the 
value of $Q^*_D$ can vary by up to a factor of three between strong crystalline and weak aggregates. However, 
they did not characterize the outcome of the disruptions at $Q^*_D$ in terms of fragment size and 
velocity distributions and their model was not appropriate to address the case of porous materials.

In our study, we use an improved version of the SPH hydrocode of Benz and Asphaug (1999), which 
now includes a model adapted to porous materials (Jutzi et al. 2008) validated at small scales by 
 a successful test against laboratory experiments using pumice targets (Jutzi et al. 2009a). Then, 
in order to explicitly characterize the fragments produced by
gravitational reaccumulation in the gravity regime, we combine it with the gravitational
$N$-body code {\it pkdgrav}, as was done to reproduce asteroid families (e.g. Michel et al. 2001, 2003, 
Jutzi et al. 2009b). We next look for the catastrophic specific energy threshold
as a function of target diameter in both the strength and gravity regime, and we determine in
more detail in the gravity regime 
the outcome properties of the disruption as a function of the model used to characterize the target.  

By explicitly accounting for the two main processes involved in
large-scale collisions (fragmentation and gravitational interaction), our
method allows us to determine not only the value of $Q^*_D$ but also the full
size and ejection velocity distributions of fragments down to the
resolution limit imposed by the numerical techniques. Moreover, we can
model different kinds of internal structure of the target and analyze
the dependency of the value of $Q^*_D$ and the fragments' properties on
the target's properties. This is an important aspect since, as described
for instance by Holsapple et al. (2002) and Asphaug et al. (2002), researchers
developing collisional evolution codes continue to debate over which values of $Q^*_D$
are appropriate for particular material properties, internal
structures, impact speeds and object diameters. In particular,
the effect of the internal structure on the outcome and on the impact
energy required for disruption is a crucial information, as it
constrains the collisional lifetime of the small body under consideration. It
also has many implications in the framework of impact risk assessment
and mitigation strategies. Indeed, it is important to make sure that a planned deflection 
does not lead to a disruption, and therefore the energy threshold for disruption is important 
information.

Here, thanks to the implementation of a model of fragmentation of porous materials, 
we consider two kinds of parent bodies, either non-porous or porous, which are generally believed to 
represent bright and dark asteroids, respectively. 

In the following, we present the results of our investigations, starting in
Section 2 by a description of the two kinds of target model for which we
will provide the specific impact energy threshold for disruption. The numerical method is then 
briefly detailed in Section 3. The catastrophic disruption energy threshold of non-porous and porous 
targets as a function of projectile's velocity and target strengths is investigated in Section 4. 
The outcomes in terms of fragment size and speed distributions
for both models are described in Section 5 and 6,
respectively. Section 7 presents a comparison between these results
allowing us to assess the sensitivity of the outcome properties and
impact energy values upon the internal structure. Conclusions and
perspectives are presented in Section 8.

\section{Target models}

The catastrophic energy threshold and corresponding fragment properties were 
determined for two basic models of the target's internal
structure, namely non-porous and porous. 

The first model consists of a purely monolithic non-porous target that 
initially contains only a distribution of incipient flaws and no damaged zones or
macroscopic voids. Its fragmentation is computed with
our SPH hydrocode (Benz and Asphaug 1994) based on a model of brittle failure of solid materials validated 
at small scale by comparison with laboratory experiments on basalt targets (Nakamura and Fujiwara 1991). 

The second model is a porous target that consists of a body containing sub-resolution pores whose sizes are 
smaller than the thickness of the shock front. The fragmentation is computed using our recent model of fragmentation of 
porous material implemented in the SPH hydrocode and validated at small scale by comparison with  
laboratory experiments on pumice targets (Jutzi et al. 2008, 2009a).

The values of $Q^*_D$ have been computed using nominal values of material parameters of basalt (non-porous targets) and pumice (porous targets). 
In order to investigate the influence of the tensile and shear strengths, we computed $Q^*_D$ for two 
additional kinds of targets by, respectively, decreasing and increasing the tensile and 
shear strengths (i.e., the von Mises yield strength $Y$) from the nominal values. These targets are defined as weak non-porous and strong porous targets.

Then we investigated the outcome in terms of fragment size and speed distributions at 
$Q^*_D$ using the nominal material properties of basalt for the non-porous targets 
and pumice for the porous ones. Other material properties will be investigated in the future to determine whether our 
results can be generalized.

\section{Numerical method}

In order to characterize $Q^*_D$ and outcome properties, we use a method and numerical codes 
based on the ones that have already allowed us to simulate successfully the formation of major 
bright-type asteroid families in different impact energy regimes (Michel et al., 2001, 2002, 2003, 2004a, 
b). More precisely, our method consists of dividing the process into two phases: a fragmentation phase 
computed by a 3D SPH hydrocode (Benz and Asphaug 1994, Jutzi et al. 2008), and a gravitational phase 
computed by the gravitational N-body code {\it pkdgrav} (Richardson et al. 2000) during which fragments 
can interact with each other due to their mutual attractions. In the strength regime at small scale, only the 
fragmentation phase needs to be computed. 

Our hydrocode was originally limited to addressing the fragmentation of brittle non-porous materials. 
A model of fragmentation of porous bodies (that accounts for the crushing of pores in addition to the damage caused by the activation of cracks) has been developed and tested recently at laboratory scale (Jutzi et al. 2008, 2009a). 
In the following we give a short overview of our method and codes, and then present our simulations. 

\subsection{Numerical model of fragmentation}

\subsubsection{Classical model of brittle failure}

To compute the fragmentation phase of the collision, we use a smoothed particle hydrodynamics (SPH) code. 
The standard gas dynamics SPH approach was extended by Benz and Asphaug (1994, 1995) to include an 
elastic-perfectly plastic material description (see, e.g., Libersky and Petschek 1991) and a model of 
brittle failure based on the one of Grady and Kipp (1980). The so-called Tillotson equation of 
state for basalt (Tillotson 1962) is used to relate the pressure to density and internal energy. We refer 
the reader to the papers by Benz and Asphaug (1994, 1995) for a detailed description of this code. This code was 
then used by Benz and Asphaug (1999) to make a first complete characterization of $Q^*_D$ for 
basalt and ice targets at different impact speeds. 

\subsubsection{Model including porosity}

Recently, our SPH impact code was extended to include a model adapted for 
porous materials (Jutzi et al., 2008, 2009a). Before presenting its main principles, we first define what 
is meant here by porosity. The scale of porosity must be defined in comparison with the other relevant 
dimensions involved in the problem, such as the size of the projectile and/or crater, etc. In particular, 
we define microscopic porosity as a type of porosity characterized by pores sufficiently small 
that their distribution can be assumed uniform and isotropic over these relevant scales. Specifically, the 
sizes of the pores are in this case smaller than the thickness of the shock front. In this paper, a 
{\it porous parent body} is considered to contain this kind of microporosity. Macroscopic porosity 
on the other hand is characterized by pores with sizes such that the medium can no longer be assumed to 
have homogeneous and isotropic characteristics over the scales of interest. In this case, pores have to be 
modeled explicitly and the hydrocode as described previously, which includes a model for non-porous brittle 
solids, can still be used. The presence of these large macroscopic voids will only affect the transfer 
efficiency and the geometry of the shock wave resulting from the impact, which can be computed 
using the existing code. This was done by Michel et al. (2003, 2004a) to model the disruption of 
pre-shattered parent bodies of S-type families. On the other hand, a body containing microporosity 
may be crushable: cratering on a microporous asteroid might be an event involving compaction rather 
than ejection (Housen et al. 1999a). Thus, for an impact 
into a microporous material, a part of the kinetic energy is dissipated by compaction which leads to less 
ejected mass and lower speeds of the ejected material. These effects cannot be reproduced by hydrocodes 
developed for the modeling of non-porous solids.

Our model is based on the so-called $P-alpha$ model 
initially proposed by Herrmann (1969) and later modified by Carroll and Holt (1972). A detailed 
description of the model and its implementation in our SPH hydrocode can be found in Jutzi et al. (2008). 

The original idea at the origin of the $P-alpha$ model is based on the separation of the volume change 
in a porous material into two parts: the pore collapse on one hand and the compression of the material 
composing the matrix on the other hand. This separation can be achieved by introducing the so-called 
distention parameter $\alpha$ defined as 
\begin{equation}
\alpha=\frac{\rho_s}{\rho}
\end{equation}
where $\rho$ is the density of the porous material and $\rho_s$ is the density of the corresponding 
solid (matrix) material. Distention can be converted to porosity using the relation porosity = $(1 - 1/\alpha)$. 

The distention parameter $\alpha$ is then used in the computation of the pressure and the deviatoric 
stress tensor. As material parameters we use those derived from our successful validation 
of the model by comparison with laboratory impact experiments into porous pumice (Jutzi et al., 2009a).

\subsection{Numerical model of the gravitational phase}\label{sec:grav}

Once the collision is over and fracture ceases, the hydrodynamical 
simulations are stopped and intact fragments (if any) are identified. These 
fragments, as well as single particles and their corresponding velocity distribution,  
are fed into an $N$-body code that computes the dynamical evolution 
of the system to late time. Note that since the total mass is fixed, the extent 
of the reaccumulation is entirely determined by the velocity field imposed by the 
collisional physics upon the individual fragments. 

Since we are dealing with a fairly large number of bodies that we want to 
follow over long periods of time, we use a parallel $N$-body 
hierarchical tree code (Richardson et al. 2000). The tree component of the code provides a convenient means
of consolidating forces exerted by distant particles, reducing the
computational cost. The parallel component divides the work evenly
among available processors, adjusting the load each timestep
according to the amount of work done in the previous force
calculation. The code uses a straightforward second-order leapfrog
scheme for the integration and computes gravity moments from tree
cells to hexadecapole order. Particles are considered to be finite-sized hard 
spheres and collisions are identified at each step using 
a fast neighbor-search algorithm. The code then 
detects and treats collisions and mergers between particles on the
basis of different options that were investigated by Michel {\it et
al.} (2002) for monolithic non-porous parent bodies. Here we use the most
realistic treatment in which a criterion based on relative speed and
angular momentum is applied: fragments are allowed to merge only if
their relative speed is smaller than their mutual escape speed and the
resulting spin of the merged fragment is smaller than the threshold
value for rotational fission.  When two particles merge, they are
replaced by a single spherical particle with the same momentum.
Non-merging collisions are modeled as bounces between hard spheres
whose post-collision velocities are determined by the amount of
dissipation occurring during the collisions. The latter is determined
in our simulations by the coefficients of restitution in the
tangential and normal directions of the velocity vectors relative to
the point of contact (see Richardson 1994 for details). The values of
these coefficients are poorly constrained; we chose to set the normal coefficient 
of restitution to $0.3$ for porous targets and $0.5$ for non-porous ones, and the tangential coefficient to $1$ (representing 
no surface friction). 
Michel et al. (2002) already found that values of the normal coefficient of restitution in the range $0.5$-$0.8$ 
led to similar outcomes for non-porous bodies, and we checked that the same holds true in the case of porous bodies 
for values in the range $0.3$-$0.5$. 

Note that Richardson et al. (2009) have improved these simulations by adding a model for 
rigid aggregates, which allows particles to stick (or bounce) at contact and thus, grow 
aggregates of different shapes. Although this improvement allows a determination of shape and 
spins of fragments, it involves a large additional computational effort which goes beyond the scope of the present paper.

\section{Simulations at the catastrophic disruption energy threshold}

To characterize $Q^*_D$ for a given impact speed, we have proceeded by trial and
errors, by making several simulations and then looking for those that led to 
a largest fragment containing $50 \%$ of the parent body's mass. We use about $2 \times 10^5$ 
SPH particles to perform the simulations. In the gravity regime, in a first
step, we computed explicitly the fragmentation phase by using the SPH
hydrocode. We obtained a first estimate of the largest fragment mass by using the iterative
procedure based on energy balance developed by Benz and Asphaug (1999).
Once the impact conditions expected to lead to the appropriate remnant mass were identified, the complete
simulation was carried out, this time including both the fragmentation and
gravitational phases. The gravitational phase was computed using a stepsize of $50$ s (several runs were 
made with a stepsize of $5$ s and obtained similar results) and was carried out 
to a simulated time of $11.6$ days after which the outcome does  
not change anymore. Note that the mass of the largest remnant found by the complete simulation (including the gravitational phase) is in a good agreement with the mass found by the iterative energy balance method (the difference is typically of the order 3-5 \%).

\subsection{Catastrophic impact energy threshold as a function of target diameter for non-porous and porous materials}\label{sec:qdd}

Table \ref{t:nommatprop} gives the nominal material properties of 
basalt and pumice used in our two models of internal structure (non-porous and porous). We define them as nominal, 
because in order to characterize the influence of the tensile and shear strengths, we will modify the values of these parameters 
to compute $Q^*_D$ in the 
following subsection.  
Table \ref{t:nomic} gives the impact conditions of our simulations (which, in the gravity regime, include the explicit computation of the gravitational phase)  for all the investigated diameters and for the two nominal models.

Note that the largest fragment from these simulations does not contain exactly
$50 \%$ of the target's mass since it would be computationally unreasonable
to look for the impact conditions that lead to the exact
value. To determine  $Q^*_D$, we performed at least 
one additional impact simulation around the specific energy threshold and we interpolated through the corresponding values of the impact energy to 
derive the exact value of $Q^*_D$. Note that in the gravity regime, we used the mass of the largest remnant obtained by the energy balance method to perform the interpolation. In all simulations presented here, we obtained $0.35 < M_{lr}/M_{pb} < 0.65$, where
$M_{lr}$ and $M_{pb}$ are the largest remnant and parent body masses,
respectively.

\begin{figure}
\centerline{\psfig{file=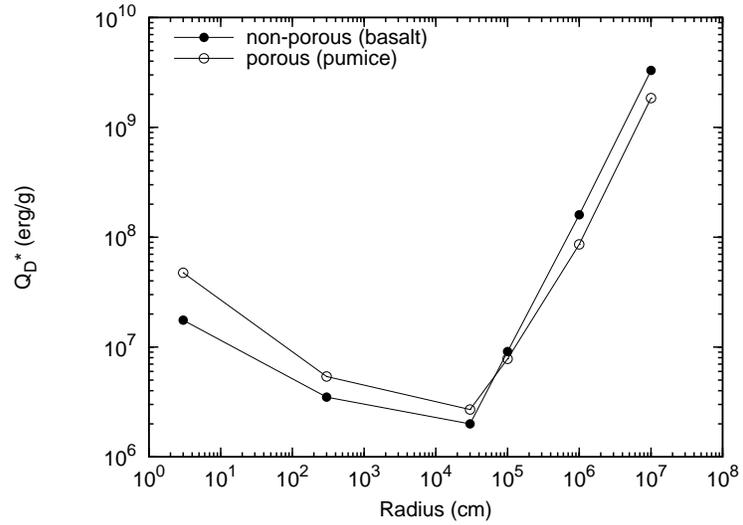,height=7cm}}
\caption{\small Catastrophic
  specific impact energy threshold $Q^*_D$ (erg$/$g) as a function of the target
  radius $R$ (cm). The impact speed and angle are $3$
  km$/$s and $45^\circ$, respectively. The internal structure of the
  target is either porous or non-porous, as indicated on the plot.  }
\label{f:qcritpornonpor3kms}
\end{figure}

Figure \ref{f:qcritpornonpor3kms} presents the relationship
between $Q^*_D$ and target diameter for the two kinds of parent
bodies and for a projectile speed of 3 km$/$s and impact angle of $45^\circ$ (which corresponds to the most probable impact angle).
As shown by Benz and Asphaug (1999) for basalt and ice targets, $Q^*_D$ increases with increasing impact angle (and decreases with decreasing impact angle). In other words, everything else being equal, a larger projectile is needed at higher impact angle to achieve the same degree of mass loss. The same holds true for both the porous and non-porous targets investigated in this paper.
The material properties of the porous target are those that provided the best match to the impact 
experiments on pumice targets (Jutzi et al. 2009a). Non-porous targets are characterized by material 
properties of basalt (Benz and Asphaug 1994), so in this case our simulations, using an improved 
version of the SPH hydrocode, revisit the 
values of $Q^*_D$ estimated by Benz and Asphaug (1999), who used similar targets.

As expected, in the strength regime, the value of $Q^*_D$
decreases with target diameter, while in the gravity regime, $Q^*_D$ increases 
with target diameter due to the gravitational attraction that
has to be overcome and that increases with the size of the target. 
However, in the strength regime (radius smaller than a few hundreds meters), porous targets are stronger than 
non-porous ones, as more energy is required to disrupt them (as found in laboratory experiments).

\begin{figure}
\centerline{\psfig{file=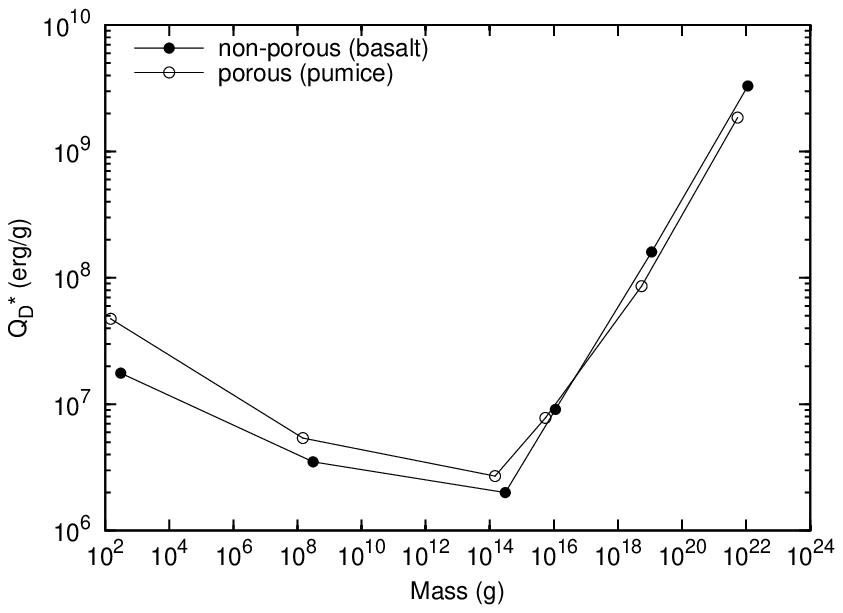,height=7cm}}
\caption{\small Catastrophic
  specific impact energy threshold $Q^*_D$ (erg$/$g) as a function of the target
  mass (g). The impact speed and angle are $3$
  km$/$s and $45^\circ$, respectively. The internal structure of the
  target is either porous or non-porous, as indicated on the plot.  }
\label{f:qcritpornonporm3kms}
\end{figure}

The opposite is true in the gravity regime, where the porous targets become weaker than non-porous ones. To explain this change of impact response we first point out that in the strength regime, the largest remnant is one large intact fragment while in the gravity regime, targets are first totally shattered by the fragmentation before building up the largest remnant through gravitational reaccumulation. One reason for the change in impact response is therefore linked to the fact that the target density of the porous target is smaller than that of non-porous targets, leading to a less efficient reaccumulation in the porous case.  We use a bulk density of $1.3$ g$/$cm$^3$ for porous targets, and $2.7$ g$/$cm$^3$ for non-porous ones, in order to be consistent with the estimated densities of dark-type and bright-type asteroids (e.g. Yeomans et al. 1997, Wilkison et al. 
2002). Therefore, target's masses are different for the two kinds of bodies at a given diameter. Figure \ref{f:qcritpornonporm3kms} shows the relation between $Q^*_D$ and mass. Compared to the relation involving sizes, the difference is not huge, but the curve for porous bodies in the gravity regime is indeed shifted toward the 
one for non-porous bodies. Note that the bulk density of the projectiles impacting both kinds of targets is set to $2.7$ g$/$cm$^3$. We checked in a few cases that using a projectile with a bulk density of $1.3$ g$/$cm$^3$ does not lead to significantly different results than those obtained with a higher density. 
Another reason for the change of impact response in the gravity regime is related to the shear strength which is generally higher in non-porous targets than in porous ones. Note that while that the value of the yield strength ($Y$) corresponds to the shear strength in non-porous materials, this is not the case in our porous material. As described in Jutzi et al. (2008), $Y$ is the yield strength of the matrix material and does not correspond to the ''bulk'' shear strength of the porous material  (which is generally lower).  
As Leinhardt and Stewart (2008) showed, the shock wave decays more rapidly in strong (high shear strength) materials than in weak  (low shear strength) materials. The authors conclude that  the stronger the material, the more energy is partitioned into overcoming the shear strength. Since our porous material has a much smaller bulk shear strength than the non-porous material, more energy is partitioned into plastic deformation in the latter case. This effect could, at least partially, compensate the effect of the dissipation of energy by compression ($P dV$ work, where $P$ is the pressure and $V$ is the volume) in porous targets. In section \ref{sec:strength} we investigate  $Q^*_D$ for different values of the shear strength. We find that in the gravity regime, $Q^*_D$ for non-porous targets  decreases significantly with decreasing shear strength which is not the case for porous targets.  
The influence of the shear strength is less strong for smaller targets (in the strength regime) since in this regime, the tensile strength dominates the outcome (size of intact fragments) and, in addition, the impact energies are lower and the resulting shock waves are less strong. 

We will investigate the fraction of the incoming energy which goes into dissipation by compaction, plastic deformation or kinetic energy of the target in a further study.

\subsection{Influence of the impact velocity}

\begin{figure}
\centerline{\psfig{file=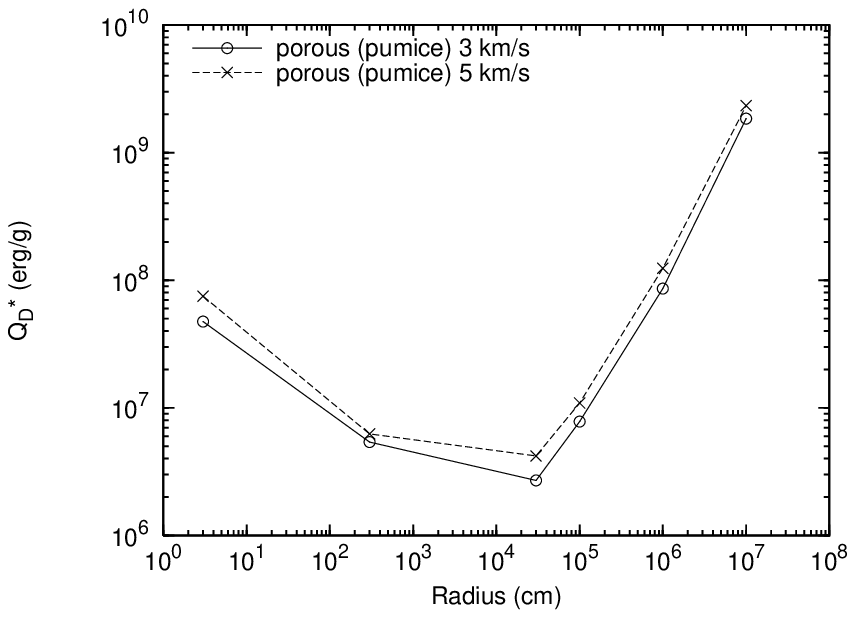,height=7cm}}
\caption{\small Catastrophic
  specific impact energy threshold $Q^*_D$ (erg$/$g) as a function of porous target
  radius $R$ (cm) for impact speeds of 3 and 5 km$/$s and an impact angle of $45^\circ$.}
\label{f:qcritporv3v5kms}
\end{figure}

\begin{figure}
\centerline{\psfig{file=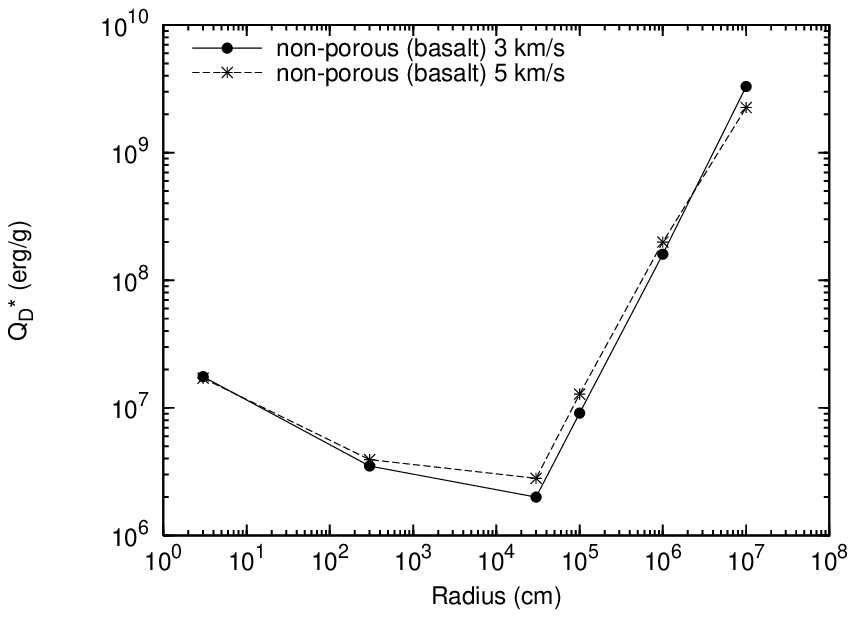,height=7cm}}
\caption{\small Catastrophic
  specific impact energy threshold $Q^*_D$ (erg$/$g) as a function of non-porous target
  radius $R$ (cm) for impact speeds of 3 and 5 km$/$s and an impact angle of $45^\circ$.}
\label{f:qcritnonporv3v5kms}
\end{figure}

The characterization of $Q^*_D$ for targets of a given size involves both the size and the speed of the projectile. For 
a given impact speed, the projectile's size is varied, which is what was done to produce Figs.~\ref{f:qcritpornonpor3kms} 
and \ref{f:qcritpornonporm3kms}. It is also important 
to determine the dependency of $Q^*_D$ on the impact speed. Figures \ref{f:qcritporv3v5kms} and \ref{f:qcritnonporv3v5kms} 
show the results for the 
the porous and non-porous targets, respectively, for impact speeds of 3 and 5 km$/$s in the gravity 
regime. In the porous case, $Q^*_D$ is systematically higher using higher impact speed. In the non-porous case, we 
find that for the largest considered target size (100 km), $Q^*_D$ becomes smaller for the higher impact speed. 
This change was already apparent in the results 
by Benz and Asphaug (1999). Stewart and Leinhardt (2008) also noticed that the slope of the $Q^*_D$ curve at $3$ km$/$s 
becomes shallower for basalt targets with large sizes and interpreted this behavior as 
a signature of a larger contribution of the projectile mass as it becomes a significant fraction of the 
target mass. However, we checked that including the projectile particles in the computation of the 
gravitational phase of the disruption of our $100$ km-radius basalt target at both impact speeds, i.e. $3$ km$/$s and 
$5$ km$/$s did not lead to a different outcome, i.e. $Q^*_D$ and the fragment size distributions are identical, and therefore 
the explanation of this change must reside elsewhere. 

We also characterized $Q^*_D$ for porous targets of 300 meters in radius using higher projectile speeds (i.e. $7$ and $10$ km$/$s, 
in addition to $3$ and $5$ km$/$s) 
and found that $Q^*_D$ keeps increasing with the impact speed, which is consistent with the trend proposed by scaling 
laws (Housen and Holsapple 1990):
\begin{equation}
Q^*_D= C \rho R^{3 \mu}V^{-3\mu+2}
\end{equation}
where $C$ is a constant, $\rho$ is the density, $V$  is the impact velocity, $R$ is the target radius and $\mu$ is the so-called 
coupling parameter. Despite the fact that the considered porous target's radius (300 meters) is close to the transition between 
the strength and gravity regimes, we can still reasonably fit this scaling relationship, which is only appropriate for the gravity regime,  
with $\mu$ = 0.43  $\pm$ 0.01 and $C$ = 4.6 $\pm$ 0.6 $\times$ 10$^{-4}$. Note that our value of $\mu$ is consistent with the values found in experiments involving porous materials (e.g. Housen and Holsapple, 1999b).
\subsection{Power law scaling}
For practical use by collisional evolution models, we fit (by eye) the Q$^*_D$ curves of 
both kinds of targets, for the impact speeds investigated, by power laws of the form: 
\begin{equation}
 Q^*_D = Q_0\left(\frac{R_{pb}}{1 cm}\right)^a + B\rho \left(\frac{R_{pb}}{1 cm} \right)^b
\end{equation}
where $R_{pb}$ is the radius of the parent body, $\rho$ its density in g/cm$^3$ and $Q_0$, $B$, $a$, $b$ are constants to be determined.
Such a functional form is often used in scaling law approaches with the two terms 
representing the strength and gravity regimes, respectively. 

\begin{figure}
\centerline{\psfig{file=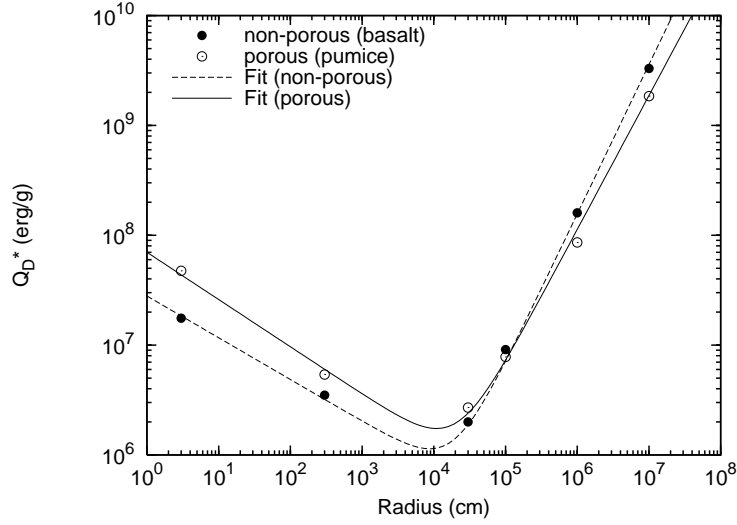,height=7cm}}
\caption{\small Fits of the catastrophic
  specific impact energy threshold $Q^*_D$ (erg$/$g) as a function of target
  radius $R$ (cm) using power laws. The impact speed is 3 km$/$s and the impact angle is $45^\circ$.}
\label{f:qcritpowerlaw3kms}
\end{figure}

Figure \ref{f:qcritpowerlaw3kms} shows our fits for the impact speed of $3$ km$/$s. The same kind of qualitative fit is 
at the origin of the values\footnote{Note that we find slightly different values for the non-porous targets than the ones previously estimated by Benz and Asphaug (1999)} given in Table 3 for the impact speed of $5$ km$/$s. As it can be seen, we find slightly different slopes for the two materials and the two impact speeds, respectively.

\subsection{Comparison with different scaling variables}
Stewart and Leinhardt (2009) proposed new variables to describe catastrophic disruption. Instead of using 
the target diameter, they use the spherical radius of the combined projectile and target masses 
at a density of $1$ g$/$cm$^3$, and the critical specific impact energy is replaced by $0.5 \mu V_i^2/M_T$ where $\mu$ 
is the reduced mass, $V_i$ is the projectile speed and $M_T$ is the total mass (projectile $+$ target). Their aim was 
to remove ambiguities (over material density and projectile-to-target mass ratio) that are inherent in the traditional 
variables ($Q^*_D$ and the target diameter). Using these new variables, they found some differences with respect to 
the results plotted with the traditional ones (see Stewart and Leinhardt 2009 for details). We checked whether the 
same holds true with our simulations, and found no difference. The reason is probably that Stewart and Leinhardt 
investigated an impact speed regime for which the projectile size is comparable to the target size, hence taking 
the projectile into account in the definition of the new variables can influence the outcome. Conversely, our 
simulations involve high impact speeds, and therefore smaller projectile sizes, so that using either set of 
variables is equal.

\subsection{Influence of the tensile and shear strengths}\label{sec:strength}

The populations of asteroids and comets are certainly composed of a wide diversity of bodies and therefore, the parameter space of their 
potential material properties is probably huge. In this section, we investigate the effect of the tensile and shear strengths on the impact 
response of our non-porous and porous targets. To do so, we model our targets using basalt parameters for non-porous ones and pumice 
parameters for porous ones, but we decrease the strengths from the nominal values (Table \ref{t:nommatprop}) in the first case, and 
increase them in the second case. The new values of tensile and shear strengths are indicated in Table \ref{t:modmatprop} and we define the 
corresponding targets as weak non-porous and strong porous targets.

\begin{figure}
\centerline{\psfig{file=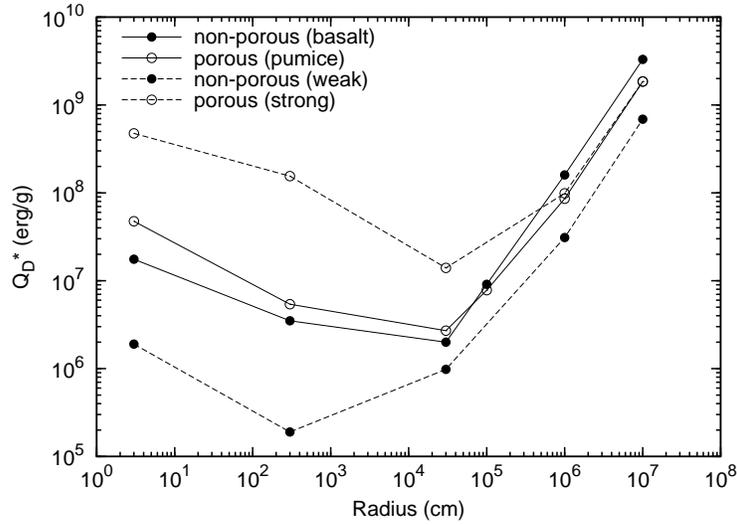,height=7cm}}
\caption{\small Catastrophic
  specific impact energy threshold $Q^*_D$ (erg$/$g) as a function of target
  radius $R$ (cm) for impact speed of 3 km$/$s and an impact angle of $45^\circ$. Targets are represented by basalt or pumice materials. The label 
weak is used for basalt non-porous targets whose tensile and shear strengths have been decreased from their nominal values, while the label  
strong is used for pumice porous targets whose tensile and shear strengths have been increased from their nominal values (see Tables \ref{t:nommatprop} and 
\ref{t:modmatprop}).}
\label{f:qcritweakstrong}
\end{figure}

As can be seen on Fig.~\ref{f:qcritweakstrong}, in the strength regime, the strong porous target is much more difficult to disrupt than the 
nominal non-porous one with identical tensile strength. On the other hand the weak non-porous target is much easier to disrupt than 
the nominal porous target which has again the same tensile strength. Therefore for the strength values investigated, non-porous targets 
are easier to disrupt in the strength regime, and the corresponding values of $Q^*_D$ can differ by several orders of magnitude. Moreover, non-porous weak targets enter in the gravity regime at smaller sizes than their 
nominal counterparts, as indicated by the earlier change 
of slope to a positive value of the $Q^*_D$ curve. The reason is that for a low enough strength, gravity starts dominating at a smaller 
size. Conversely, the slope of the $Q^*_D$ curve takes a positive value at larger sizes for the porous strong targets, because 
when the strength is high, gravity starts influencing the outcome when the size of the target is large enough to compensate from 
the high strength. 

In the gravity regime, the dependency on strength is less dramatic, especially in the porous case for which the effect of strength is almost 
meaningless. In the non-porous case, the weak target is much easier to disrupt than the nominal case, in agreement with the results of 
Leinhardt and Stewart (2008). Moreover, the values of $Q^*_D$ for these weak non-porous targets become smaller than for porous ones. 
Therefore, one cannot say that a porous target is either systematically easier or more difficult to disrupt than a non-porous one, as it 
depends on its strength properties and not only on porosity.

\section{Outcome properties for a non-porous parent body}

In this section, we analyze the outcome properties from the
disruptions of our nominal non-porous targets corresponding to the impact conditions 
described in Table \ref{t:nomic}. In this paper, we concentrate on the gravity regime only. 
Although the ratios of the largest remnant to the target mass 
obtained from our simulations do not exactly equal $0.5$, they remain close enough to $0.5$ to reasonably assume that the 
outcome properties are still a good representation of what would occur at exactly $Q^*_D$. Our aim is to
determine whether simple rules can be derived from the analysis of
these outcome properties that could be easily implemented in a
collisional evolution code. Here, we limit these properties to the
fragment size and ejection velocity distributions.

\begin{figure}
\centerline{\psfig{file=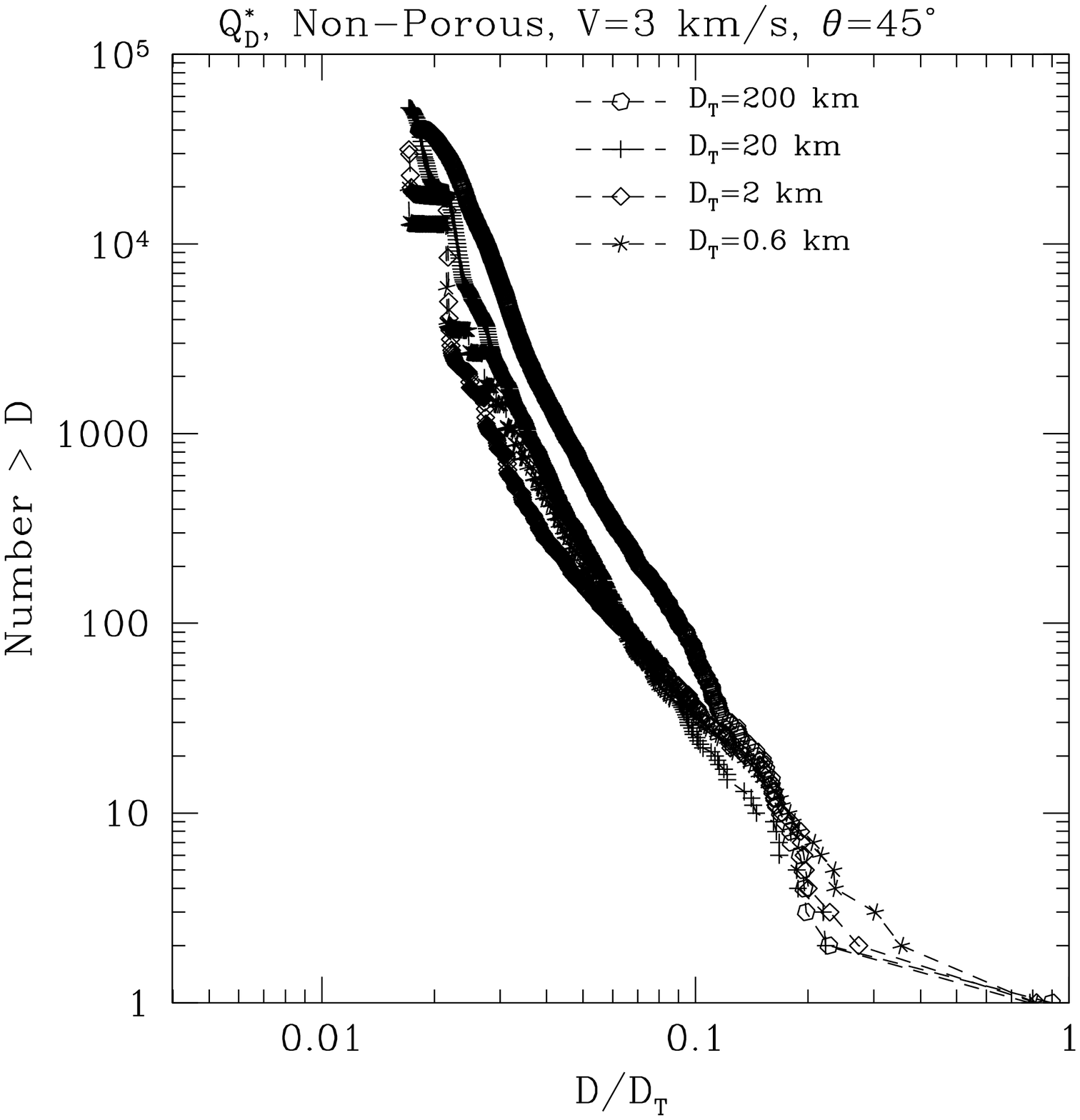,height=7cm}}
\caption{\small Cumulative diameter (km) distributions of
  the fragments of simulations at $Q^*_D$ with an impact speed $V$ of $3$ km$/$s 
and an angle of impact $\theta$ of $45^\circ$. The
  targets are non-porous and their sizes (diameter $D_T$) are indicated
  on the plot. The fragments' diameters $D$ 
  are normalized to that of the target $D_T$, for a direct comparison.} 
\label{f:cumunonporous3kms}
\end{figure}

Figure \ref{f:cumunonporous3kms} shows the fragment size distributions obtained from the
disruption of non-porous targets with different diameters. Fragment sizes 
are normalized by target diameters to allow a direct comparison. For all the
considered target's diameters, the resulting distributions look very
similar. This is a practical feature for implementation in a collisional evolution model. 

\begin{figure}
\centerline{\psfig{file=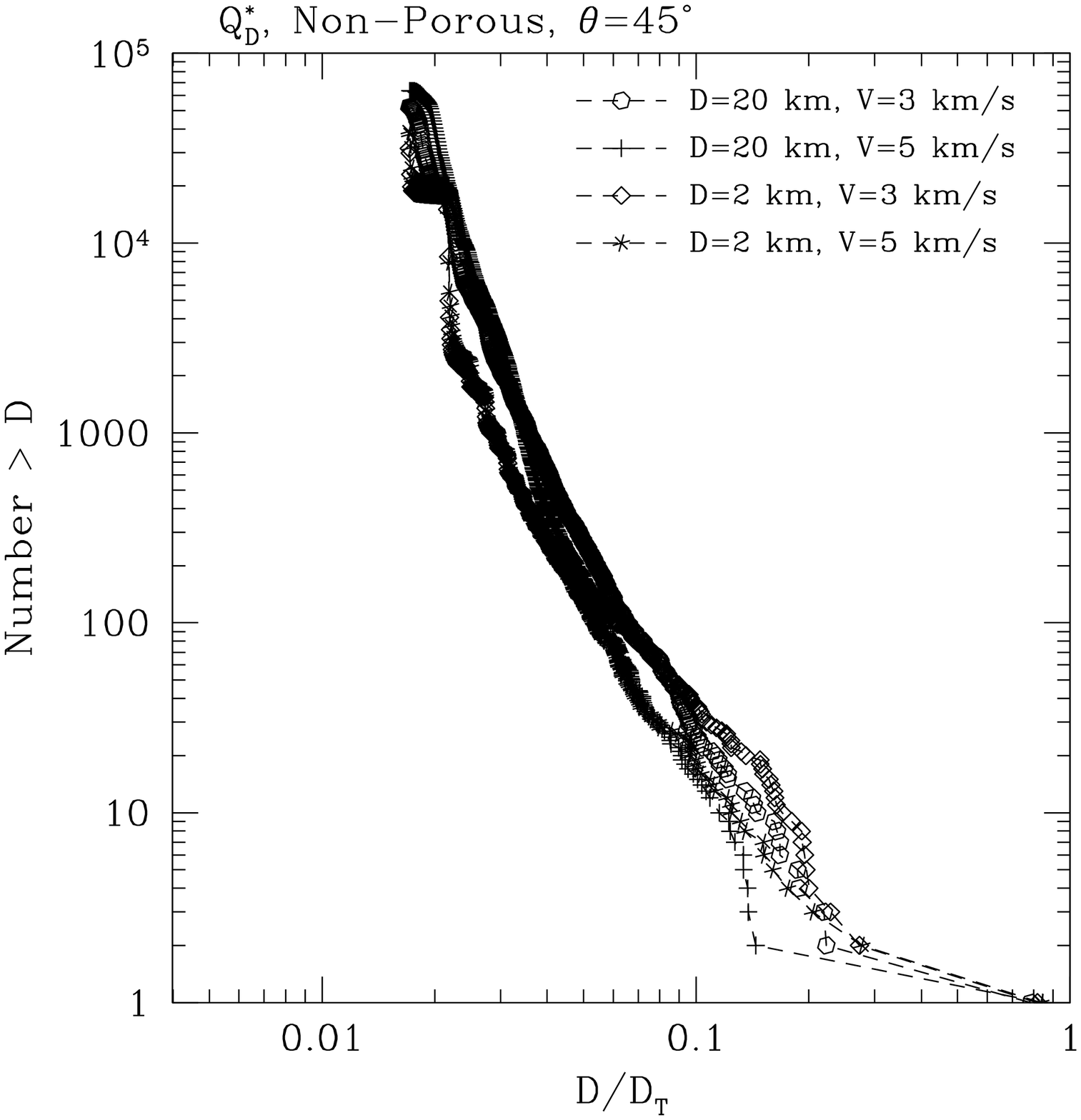,height=7cm}}
\caption{\small Cumulative diameter (km) distributions of
  the fragments of simulations at $Q^*_D$ with an impact speed $V$ of either $3$ km$/$s 
or 5 km$/$s, and an angle of impact $\theta$ of $45^\circ$. The
  targets are non-porous and their sizes (diameter $D_T$) are indicated
  on the plot. The fragments' diameters $D$ 
  are normalized to that of the target $D_T$, for a direct comparison.} 
\label{f:cumunonporV3V5kms}
\end{figure}

We also compared the fragment size distribution from a disruption at $Q^*_D$ using 
two different impact speeds. As shown on Fig.~\ref{f:cumunonporV3V5kms}, there is no large 
sensitivity on the impact speed, at least in the considered range. 

\begin{figure}
 \centerline{\psfig{file=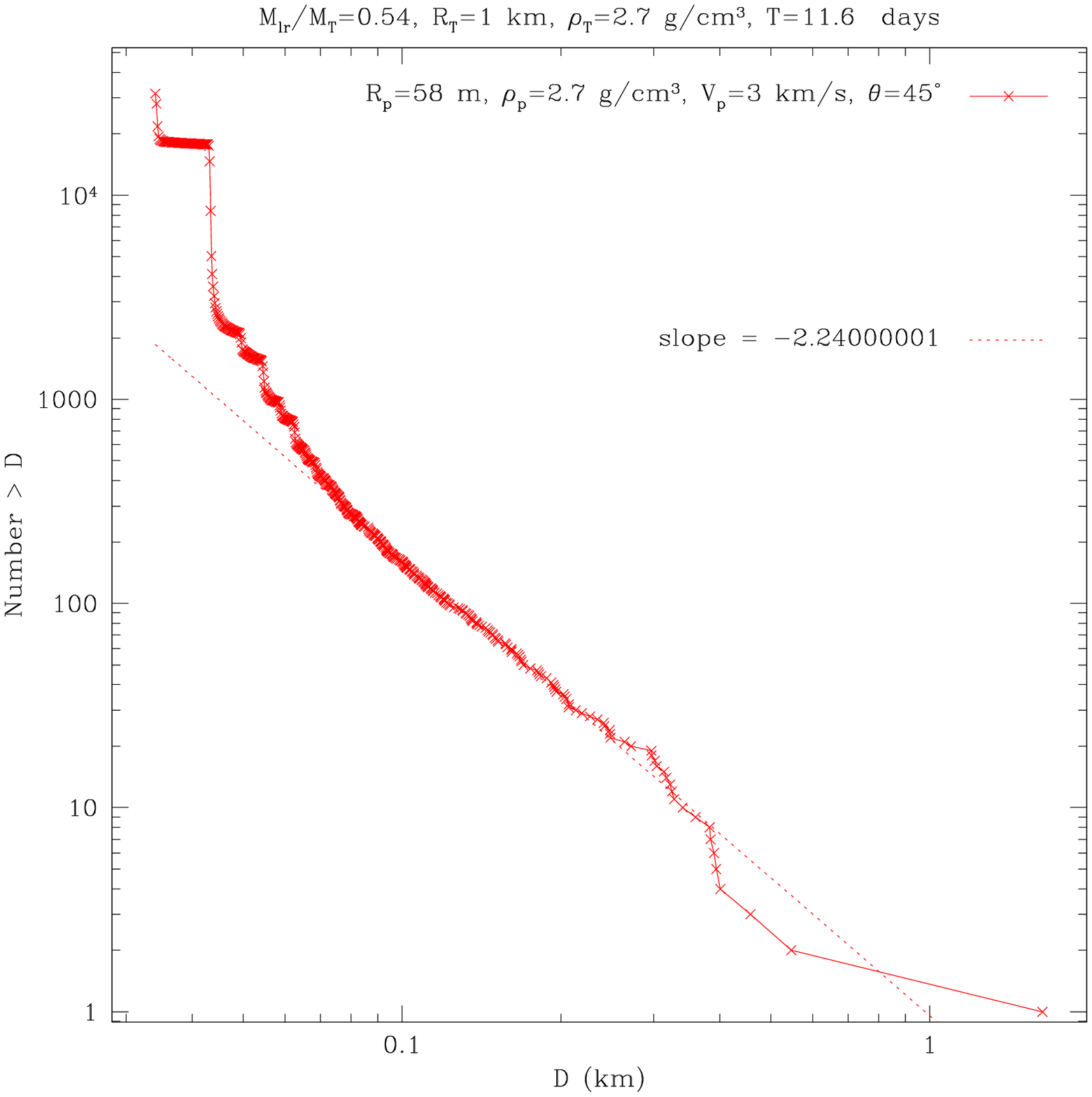,height=10cm}}
\caption{\small Cumulative diameter (km) distributions of
  the fragments of the simulation of disruption of the $2$ km-size
  non-porous target at $Q^*_D$ with an impact speed $V_p$ of $3$ km$/$s 
  and an angle of impact $\theta$ of $45^\circ$. A reasonable fit to this
  distribution with a single slope is indicated on the plot. }
\label{f:cumuslopenonpor1km3kms}
\end{figure}

Therefore, in a collisional evolution model, it can be reasonably assumed that both the shape and slope of the fragment size 
distribution from the disruption of a non-porous target at $Q^*_D$ do not depend 
on the target's diameter and impact speed, and thus can take one single form. For practical 
use by collisional evolution models, it is thus possible to characterize a reasonable fit of the 
fragment size distribution that is valid for all target diameters by a single
power law of the form $N(>D) \propto D^\alpha$ where $N(>D)$ is the
number of fragments with diameter greater than $D$ and $\alpha$ is the
power-law exponent. \Fig{cumuslopenonpor1km3kms} shows such a fit for a target
of $1$ km radius. We find that a value of the power-law exponent in the range between $-2.2$ 
and $-2.7$ can be considered to reasonably fit the size distributions for all the considered 
target diameters. 
 
\begin{figure}
\centerline{\psfig{file=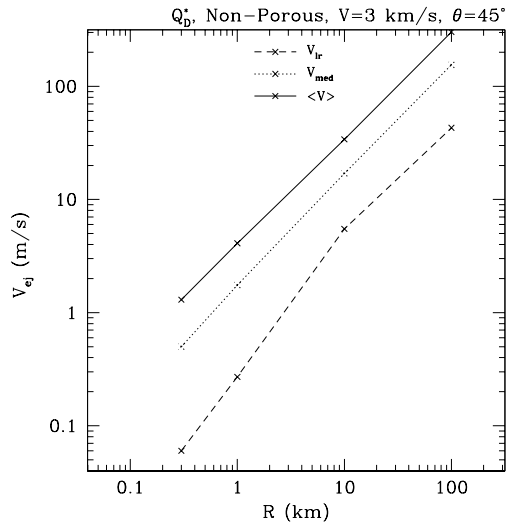,height=7cm}}
\caption{\small Different fragment ejection speeds as function of the radius $R$ 
  of non-porous targets, disrupted at $Q^*_D$ with an impact speed $V$ of 3 km$/$s and an 
impact angle $\theta$ of $45^\circ$. $V_{lr}$ stands for the largest remnant's speed, while
  $V_{med}$ and $<V>$ are the median and average fragments' speeds,
  respectively.} 
\label{f:vejnonpor3kms}
\end{figure}

\begin{figure}
\centerline{\psfig{file=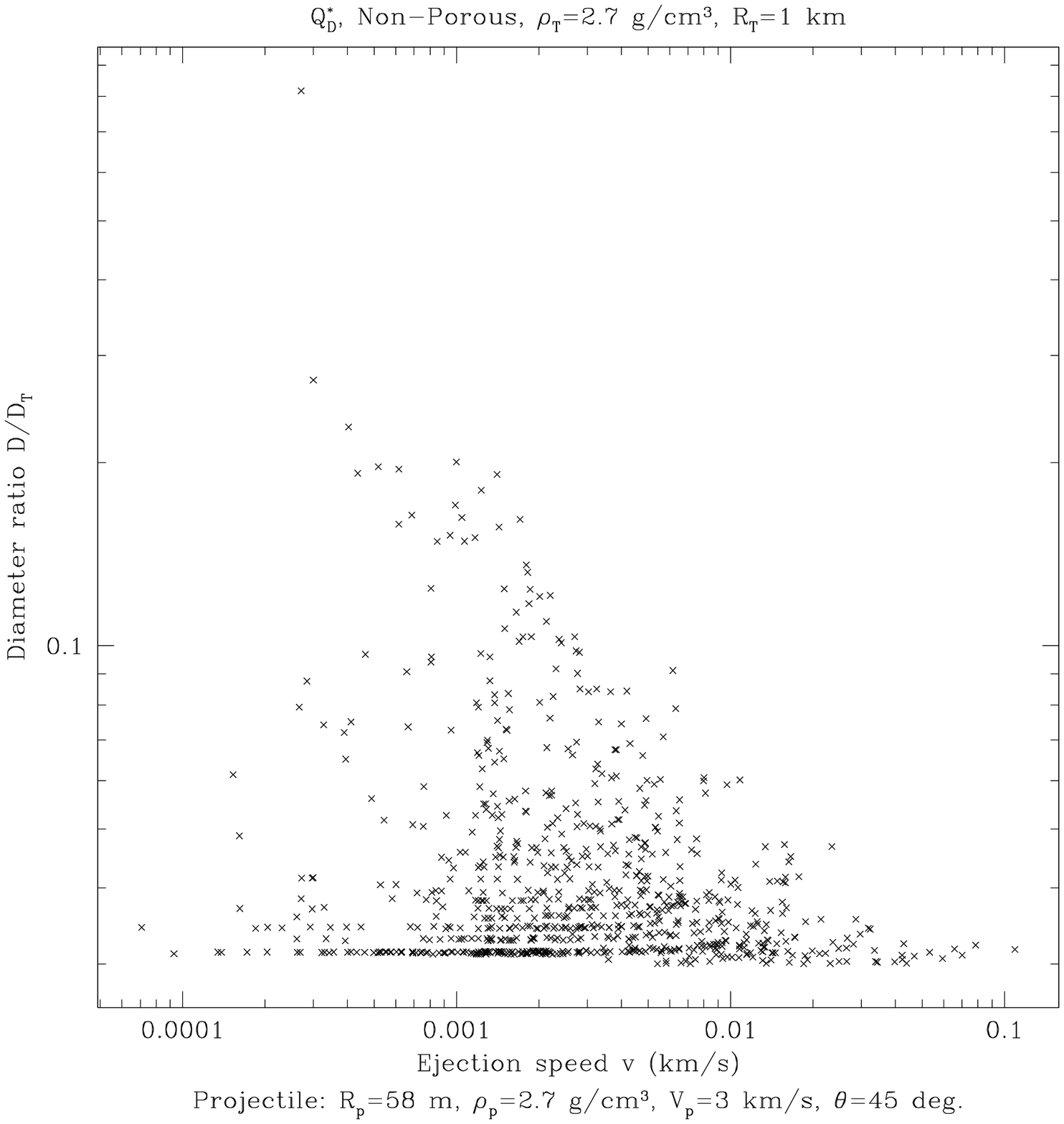,height=10cm}}
\caption{\small Fragment's diameter $D$ (normalized to that of the parent body
$D_T$) \vs ejection speed obtained from the
breakup at $Q^*_D$ of a non-porous target, $1$ km in radius. The 
impact speed $V_p$ is $3$ km$/$s and the impact angle is 
$\theta=45^\circ$.  Only fragments with size above the resolution limit
(i.e.\ those that underwent at least one reaccumulation event) are shown here. }
\label{f:sizevelnonpor1e53kms}
\end{figure}

The analysis of the fragment ejection speeds, and in particular the average,
median and largest remnant speeds, show that their values scale with the
target diameter (see Fig.~\ref{f:vejnonpor3kms}). A linear fit
applies to the relationship between either the average or median speeds and the 
target diameter in the entire diameter range represented in a log-log plot. 
Another interesting property is the relationship between fragment speed and mass 
(Fig.~\ref{f:sizevelnonpor1e53kms}). The same
result as obtained by Michel et al. (2004a) in a different impact energy
regime is found here, i.e. smaller fragments tend to have
greater ejection speeds than larger ones. However, there is still a wide
spread of values for fragments of a given mass, which makes it difficult to define
a power-law relationship between fragment masses and speeds, such as the ones
often used in collisional evolution models (see e.g., Davis 2003).

\section{Outcome properties for porous parent bodies}

The same simulations were 
performed using our nominal porous targets. The impact conditions are given in Table
\ref{t:nomic}.

\begin{figure}
\centerline{\psfig{file=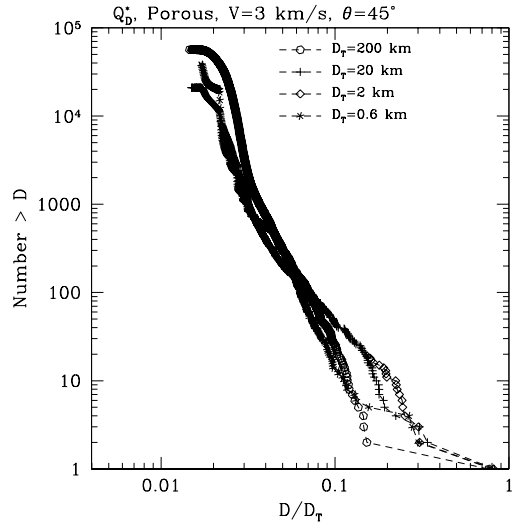,height=7cm}}
\caption{\small Same as Figure \ref{f:cumunonporous3kms} for porous parent bodies. \hspace{8cm}$ $}
\label{f:cumuporous3kms}
\end{figure}

\begin{figure}
\centerline{\psfig{file=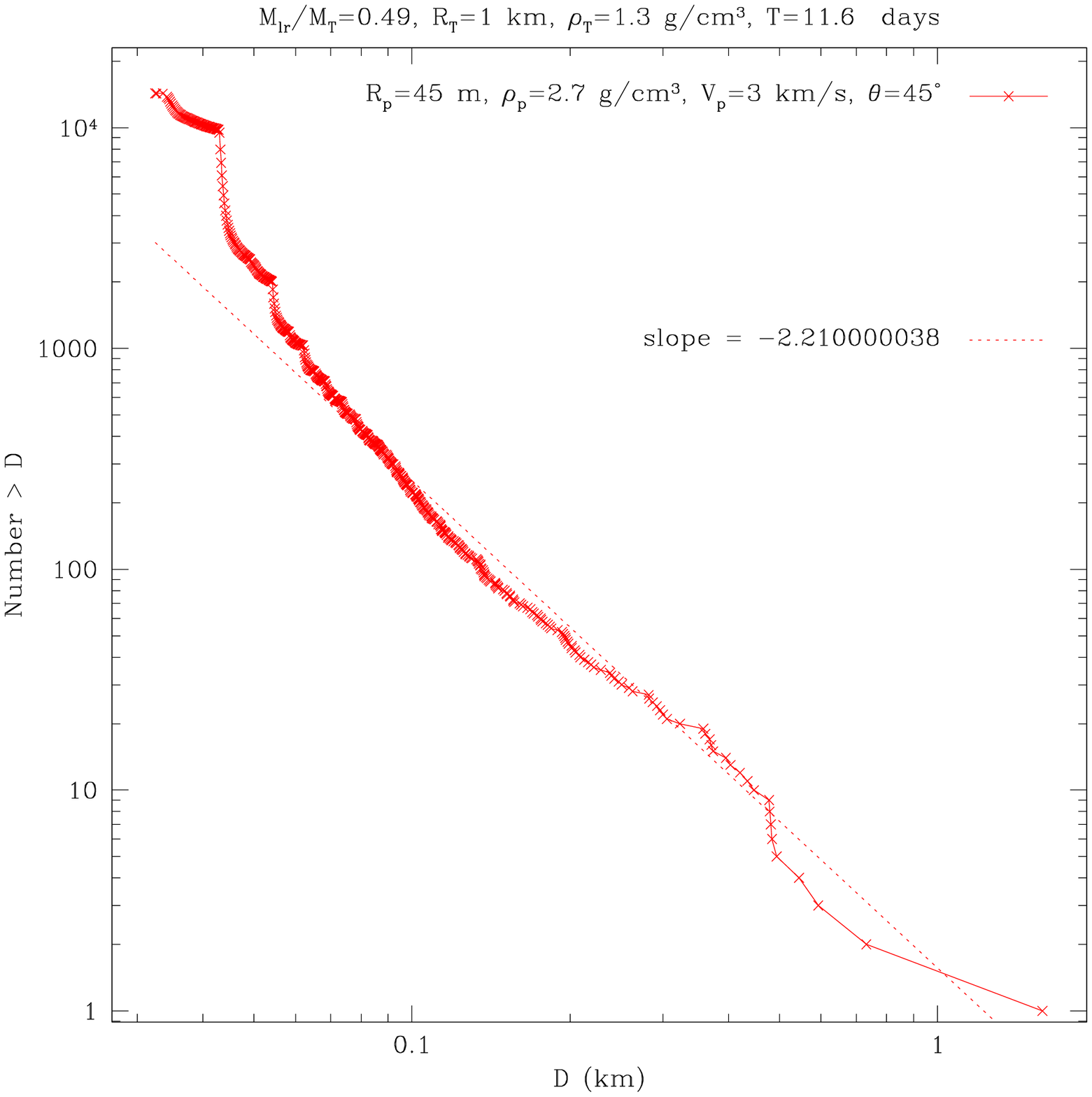,height=10cm}}
\caption{\small Cumulative diameter (km) distributions of
  the fragments of the simulation of disruption of the $2$ km-size
  porous target at $Q^*_D$ with an impact speed $V_p$ of $3$ km$/$s 
  and an angle of impact $\theta$ of $45^\circ$. A reasonable fit of this
  distribution with a single slope is indicated on the plot. }
\label{f:cumuslopepor1km3kms}
\end{figure}

For all the considered target diameters, the resulting size distributions
look very similar (see Fig.~\ref{f:cumuporous3kms}). On this plot, all the distributions 
essentially overlap with each 
other, except in a small range of sizes ($0.1<D/D_T<0.3$). Thus, as we did for non-porous targets, 
for practical use by collisional evolution models, 
we characterized a reasonable fit of the fragment
size distributions that is valid for all target diameters with a single
power law. \Fig{cumuslopepor1km3kms} shows such a fit for the target
that is $1$ km in radius. We find a best-fit value of the power-law exponent in the same range as 
for the non-porous case, between about $-2.2$ 
and $-2.7$, that can be considered to reasonably describe the size distributions for all the considered 
target diameters.

\begin{figure}
\centerline{\psfig{file=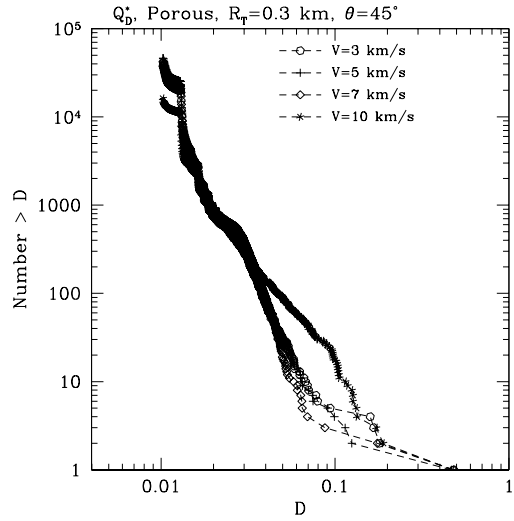,height=7cm}}
\caption{\small Cumulative diameter (km) distributions of
  the fragments of simulations at $Q^*_D$ with an impact speed $V$ of $3$, $5$, 
$7$ or $10$ km$/$s, and an angle of impact $\theta$ of $45^\circ$. The
  targets are porous and $300$ m in radius.} 
\label{f:cumuporV3V10kms}
\end{figure}

We also characterized the fragment size distribution from a disruption at $Q^*_D$ of 
a porous targets of $300$ m in radius using 
four different impact speeds. As shown in Fig.~\ref{f:cumuporV3V10kms}, we find again that there is no large 
sensitivity on the impact speed, except for the highest one (10 km$/$s), which shows some discrepancy in the diameter range between 
about $0.04$ km and  $0.2$ km. Whether this is meaningful or not will require a deeper investigation of the process at such high 
speeds, which we leave for future studies. 

\begin{figure}
\centerline{\psfig{file=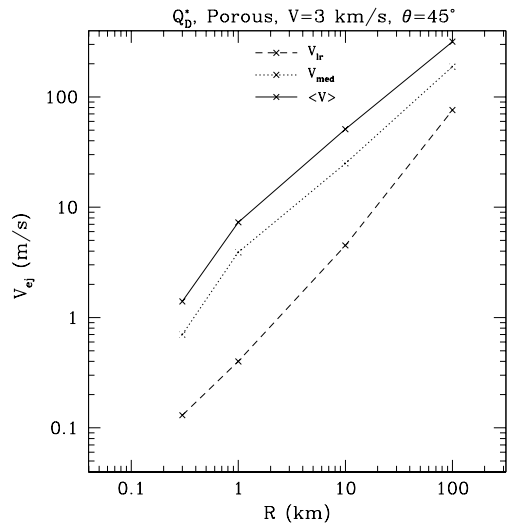,height=7cm}}
\caption{\small Same as Fig.~\ref{f:vejnonpor3kms} for the case of porous targets. \hspace{8cm}$ $}
\label{f:vejpor3kms}
\end{figure}

Concerning the ejection speeds, the
average, median and largest remnant speeds increase with increasing target 
diameter (see Fig.~\ref{f:vejpor3kms}). 

\begin{figure}
\centerline{\psfig{file=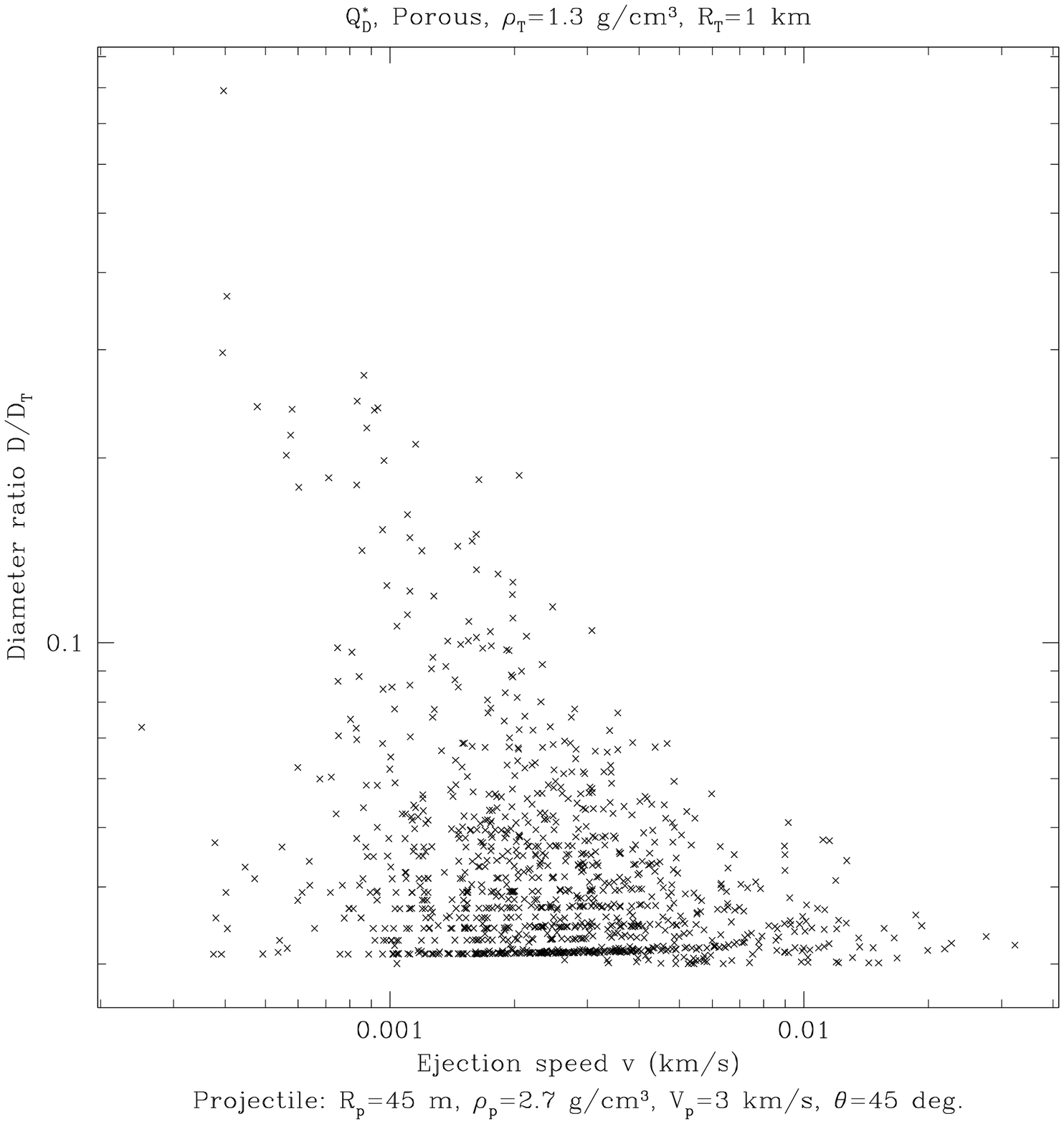,height=10cm}}
\caption{\small Fragment's diameter $D$ (normalized to that of the parent body
$D_T$) \vs ejection speed obtained from the
breakup at $Q^*_D$ of a porous target, $1$ km in radius. The 
impact speed $V_p$ is $3$ km$/$s and the impact angle is 
$\theta=45^\circ$.  Only fragments with size above the resolution limit
(i.e.\ those that underwent at least one reaccumulation event) are shown here. }
\label{f:sizevelpor1e53kms}
\end{figure}

As shown in Fig.~\ref{f:sizevelpor1e53kms},
a wide spread of ejection speeds of fragments of a given mass exists. Consequently it is again 
difficult to define a power-law relationship between fragment masses and
speeds.

\section{Comparison between the outcomes for non-porous and porous targets}

The analysis of the outcome properties from the disruption of non-porous and
porous targets has allowed us in each case to identify some systematic
behaviors, which are either independent of target size and impact speed (e.g. fragment size
distribution) or scale with the target size (e.g. median and average ejection speeds). In
this section, we analyze the systematic differences that can be identified and
that are due to the different internal structures of the parent body. Note that we consider again only the gravity regime.

\subsection{Fragment size distribution}

\begin{figure}
\centerline{\psfig{file=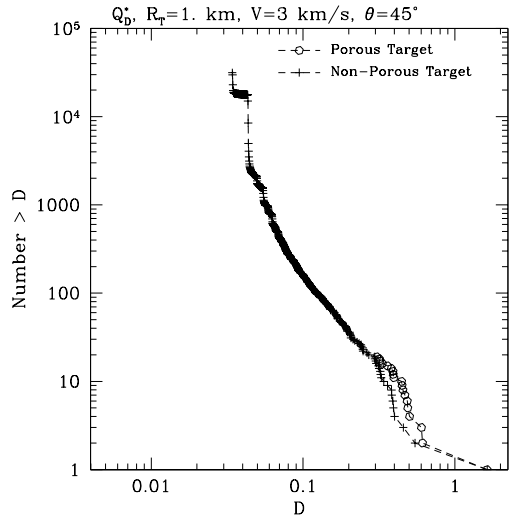,height=7cm}}
\caption{\small Comparison between the size distributions obtained from the disruption of 1 km-radius porous and non-porous targets 
with an impact speed $V$ of $3$ km$/$s and an impact angle $\theta$ of $45^\circ$.}
\label{f:cumupornonpor1e5cm3kms}
\end{figure}

For a given internal structure, the shape of the fragment size distribution certainly depends on the impact 
energy regime. For instance, in the case of a monolithic non-porous parent body, higher
impact energies lead to a more continuous fragment size distribution (Michel
et al. 2001, 2002, Durda et al. 2007). Moreover, Michel et
al. (2003) found that the size distribution of pre-shattered non-porous targets tends to be more
continuous than those of monolithic non-porous targets, for all considered
target diameters. Obviously, the parameter space could not be covered exhaustively, so it cannot be guaranteed that these 
conclusions can be generalized. Here, at impact energies close or equal to $Q^*_D$, we actually 
find that there is not much difference between the size distributions obtained 
from either a non-porous target or a porous one, at least for the two materials considered (basalt and pumice). 
For instance, Fig.~\ref{f:cumupornonpor1e5cm3kms} shows similar size distributions 
obtained from the disruption of 1 km-radius porous and non-porous targets. As we indicated 
in previous sections, the power-law exponent that can be used to fit the size distributions is in the same range for 
both kinds of targets.

\subsection{Ejection speeds}

\begin{figure}
\centerline{\psfig{file=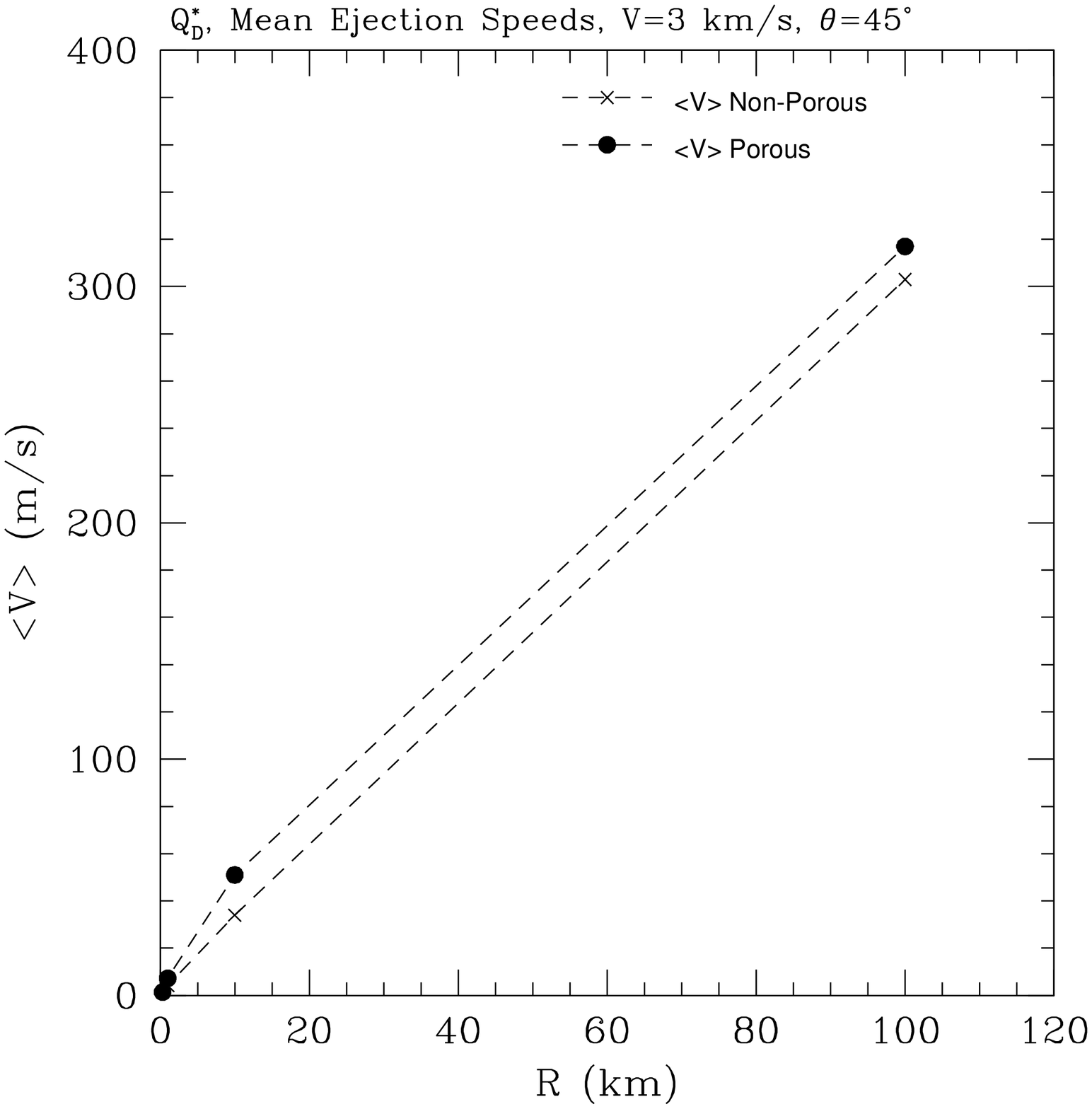,height=7cm}}
\caption{\small Mean ejection speed $<V>$ as a function of target radius $R$. 
The impact speed $V$ is $3$ km$/$s and the impact angle $\theta$ is 
$45^\circ$. The results from the two kinds of target's internal structure are shown, as
  indicated on the plot.}
\label{f:vmeanpornonpor3kms}
\end{figure}

\begin{figure}
\centerline{\psfig{file=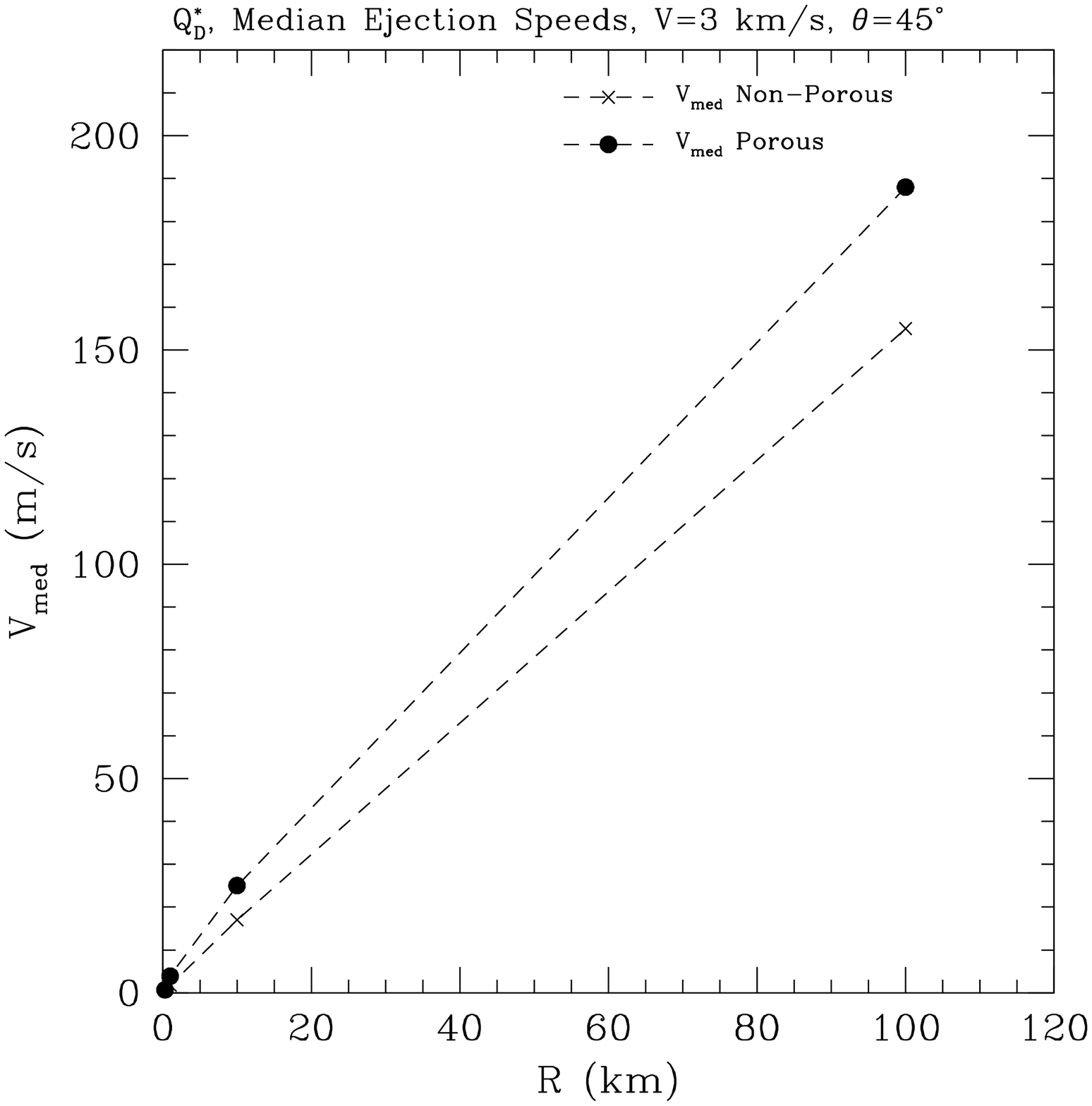,height=7cm}}
\caption{\small Median ejection speed $V_\mathrm{med}$ as a function of target 
  radius $R$. The impact speed $V$ is $3$ km$/$s and the impact angle $\theta$ is 
$45^\circ$. The results from the two kinds of target's internal structure are
  shown, as indicated on the plot.}
\label{f:vmedpornonpor3kms}
\end{figure}

The average and median speeds are slightly higher for the porous targets than for the 
non-porous ones. This can be seen in Table 5 and in Figs.~\ref{f:vmeanpornonpor3kms} 
and \ref{f:vmedpornonpor3kms}. As discussed in Section \ref{sec:qdd}, this (counter intuitive) result can probably be explained by the higher density and strength of the non-porous material. Note that these comparisons are made from 
simulations leading to values of $M_{lr}/M_{pb}$ which are not exactly the 
same in all cases, so small differences must be interpreted 
with caution. Hence, we can reasonably conclude that average and median 
speeds are of the same order in general for both kinds of target's internal 
structure, although there is a systematic trend toward higher values for the porous targets.

To easily implement these results in a collisional evolution model, one can
assume that the largest remnant, average and median ejection speeds scale with
the target's diameter. Then, slightly greater values can be assigned to 
fragments from porous targets. 
  
\section{Conclusion}

In this paper, we presented the results of complete simulations of
disruptions that allowed the determination of
the relationships between the specific impact energy threshold for disruption, 
called $Q^*_D$, and both the target diameter and its internal
structure represented by porous and non-porous materials made of pumice and basalt. We 
confirmed the results from previous studies indicating that $Q^*_D$ first decreases with 
target size in the strength regime and then increases with target size 
in the gravity regime. Moreover, we found that a porous body (as defined by our nominal model) requires more energy to be disrupted than its non-porous counterpart. In the gravity regime, the situation is reversed but the difference remains small. This might explain why, to first order, collisional evolution models could get away with only a single scaling law to reproduce the main characteristics of the asteroid populations (e.g. Bottke et al. 2005), despite the 
wide variety of internal properties that they can have. 

However, by changing the nominal values of the tensile and in particular the shear strength of our targets, we found that in the gravity regime, the value of $Q^*_D$ is not greatly influenced by the assumed 
strength for porous targets. Conversely, in the case of non-porous targets, the value of $Q^*_D$ decreases significantly with decreasing strength. Therefore, in the gravity regime one cannot reasonably assume systematically that porous targets are easier or more difficult to disrupt than non-porous ones, as it depends on the 
assumed strengths of the latter. We plan to investigate the effect of other material properties as well as other kinds of materials in the 
future. 

The value of $Q^*_D$ also depends on the impact velocity and we find similar tendency as previous studies, namely 
an increase of $Q^*_D$ with the impact speed. We then propose a scaling with speed for our 
porous targets, and find that it requires parameters that are consistent with what is expected for 
porous materials. 

We next determined the outcome properties of the disruptions at
$Q^*_D$, limited here to the fragment size and ejection velocity
distributions in the gravity regime. We found that the size distributions keep the same
qualitative aspect independent of the parent body's size and impact speed in the investigated range. For
both porous and non-porous parent bodies, they can be represented by a 
power-law whose exponent can be used to approximate the distributions 
produced from all target's sizes. The
average and median ejection speeds show also some systematic trends, i.e. 
they increase systematically with the target's diameter. Moreover,
they are of the same order for both kinds of parent bodies, although slightly higher in the porous case.

These results (although limited to a particular specific impact energy) can be easily implemented in numerical algorithms aimed at 
studying the collisional evolutions of small body populations. They
provide systematic trends in the outcome properties and scaling laws
for $Q^*_D$, at least for the two kinds of target internal
structure that we investigated, using nominal values of their material properties. Obviously, the real internal structures of small bodies
cannot be limited to these two models and we plan to develop other
models and use different material properties in order to study their resistance as well as the outcome
 properties of their disruption. While we limited our study to monolithic bodies, 
pre-shattered bodies and rubble piles, which may contain macroscopic
voids and/or some microporosity, are likely to be present in the asteroid population. 
It is then essential to understand
how macroscopic voids alone or combined with microporous properties as considered in this paper can influence the critical specific 
impact energy for disruption and whether they show some signature in
the outcome properties. Then, using an improved version of our $N$-body code (Richardson et al. 2009), we shall also be 
able to characterize the spin and shape distributions of fragments, in addition to the size and velocity distributions. 

It is already clear that a deep understanding of collisions between small
bodies requires the investigation of a huge parameter space. As a
conclusion, there is room for a great number of studies in order to
characterize the impact energies and outcome properties that will
allow us to provide to collisional evolution models the different recipes
that are valid for all impact conditions and kinds of real small bodies.  
Furthermore, this information is crucial to assess the efficiency
of mitigation techniques aimed at deflecting a potential impactor with
the Earth, as a first step requires a characterization of which impact conditions prevent the disruption of the body and 
rather permit its deflection. 

\section*{Acknowledgments}
This work was supported by the European Space Agency Advanced Concepts Team on the basis of the Ariadna contract $20782/07$ 
{\it NEO Encounter 2029}. M.J. and W.B. acknowledge support from the Swiss National Science Foundation. P.M. acknowledges the support of the French 
{\it Programme National de Plan\'etologie} (PNP), the French Program {\it Origine de Plan\`etes et de la Vie} (OPV), and the cooperation 
program CNRS-JSPS 2008-2009. D.C.R. acknowledges support from the grant NNX08AM39G (NASA). 

\section*{References}

\begin{description}

\item{}
Asphaug, E., Ryan, E.V., Zuber, M.T., 2002. Asteroid interiors. In: Bottke, W.F., Cellino, A., Paolicchi, P., Binzel, R.P. (Eds.), Asteroids III. Univ. of Arizona Press, Tucson, pp. 463-484.

\item{}
Benz, W., Asphaug, E., 1994. Impact simulations with fracture. I. Method and tests. Icarus 107, 98-116.

\item{}
Benz, W., and Asphaug, E., 1995. Simulations of brittle solids using smooth particle hydrodynamics. 
Computer Physics Communications 87, 253.

\item{}
Benz, W., and Asphaug, E., 1999. Catastrophic Disruptions Revisited. Icarus 192, 5-20.

\item{}
Bottke, W.F., Durda, D.D., Nesvorny, D.,  Jedicke, R., Morbidelli, A.,  Vokrouhlick«y, D., and Levison, H. 2005. The fossilized size 
distribution of the main asteroid belt. Icarus 175, 111.

\item{}
Carroll, M.M., Holt, A.C., 1972. Suggested modification of the $P$ - $\alpha$ model for porous materials. 
J. Appl. Phys. 43, 759-761.

\item{}
Davis, D. R. 2003. The experimental and theoretical basis for studying collisional disruption in the solar system. 
Impacts on Earth, 113.

\item{}
Durda, D.D., Bottke, W.F. , Nesvorny, D.,  Enke, B. L., Merline, W. J., Asphaug, E., Richardson, D. C.,  2007. 
Size-frequency distributions of fragments from SPH/N-body simulations of asteroid impacts: Comparison with observed asteroid 
families. Icarus 186, 498.

\item{}
Grady, D.E., Kipp, M.E., 1980. Continuum modelling of explosive 
fracture in oil shale. Int. J. Rock Mech. Min. Sci. \& 
Geomech. Abstr. 17, 147--157.

\item{}
Herrmann, W., 1969. Constitutive equation for the dynamic compaction of ductile porous materials. 
J. Appl. Phys., 40, 2490-2499.

\item{}
Holsapple, K.A., Giblin, I., Housen, K., Nakamura, A., Ryan, E., 2002. Asteroid impacts: Laboratory experiments and scaling laws. 
 In: Bottke,W.F., Cellino, A., Paolicchi, P., Binzel, R.P. (Eds.), Asteroids III. Univ. of Arizona
Press, Tucson, pp. 443.

\item{}
Housen, K. R., and Holsapple, K. A.,  1990. On the fragmentation of asteroids and planetary satellites. Icarus 84, 226-53.

\item{}
Housen, K. R., Holsapple, K. A., Voss, M. E. ,1999a. Compaction as the origin of the unusual craters
on the asteroid mathilde. Nature 402, 155157.

\item{}
Housen, K. R., Holsapple, K. A.,  1999b. Scale Effects in Strength-Dominated Collisions of Rocky Asteroids, Icarus, 142, pp. 21

\item{}
Jutzi, M., Benz, W., Michel, P., 2008.  Numerical simulations of impacts involving porous bodies. I. Implementing sub-resolution 
porosity in a 3D SPH hydrocode. Icarus 198, 242-255.

\item{}
Jutzi, M., Michel, P., Hiraoka, K.,  Nakamura, A. M., Benz W., 2009a. Numerical simulations of impacts involving porous bodies: II. 
Comparison with laboratory experiments. Icarus, 201, 802-813.

\item{}
Jutzi, M., Michel, P., Benz, W., Richardson, D.C., 2009b. The formation of the Baptistina family by catastrophic disruption: porous versus non-porous parent body. MAPS, submitted

\item{}
Leinhardt, Z. M., Stewart, S.T., 2008. Full numerical simulations of catastrophic small body collisions.
Icarus 199, 542-559.

\item{}
Libersky, L.D., Petschek, A.G., 1991. Smooth Particle Hydrodynamics with strength of materials. In: Trease, 
Fritts, Crowley (Eds.), Proc. Next Free-Lagrange Method, Lecture Notes in Physics 395, Springer-Verlag, Berlin, 
pp. 248-257.

\item{}
McGlaun, J. 1990. CTH - a three dimensional shock physics code. Int. J. Impact Eng 10. 

\item{}
Michel, P., Benz, W., Tanga, P., Richardson, D.C., 2001. Collisions
and gravitational reaccumulation: Forming asteroid families and
satellites.  Science 294, 1696-1700.

\item{}
Michel, P., Tanga, P., Benz, W., Richardson, D.C., 2002. Formation of asteroid families by catastrophic disruption: Simulations 
with fragmentation and gravitational reaccumulation. Icarus 160, 10-23.

\item{}
Michel, P., Benz, W., Richardson, D.C., 2003. Disruption of
fragmented parent bodies as the origin of asteroid
families. Nature 421, 608-611.

\item{}
Michel, P., Benz, W., Richardson, D.C., 2004a. Catastrophic disruption of pre-shattered parent bodies. Icarus 168, 420-432.

\item{}
Michel, P., Benz, W., Richardson, D.C., 2004b.
Catastrophic disruption of asteroids and family formation: a review of numerical simulations including both fragmentation and gravitational reaccumulations. PSS 52, 1109-1117.

\item{}
Nakamura, A.M.; Fujiwara, A., 1991. Velocity distribution of fragments formed in a simulated collisional disruption. Icarus 92, 132-146.

\item{}
Richardson, D.C., 1994. Tree Code Simulations of Planetary Rings. MNRAS 269, 493-511.

\item{}
Richardson, D. C., Quinn, T., Stadel, J., Lake, G., 2000. Direct large-scale N-body simulations of planetesimal dynamics. 
Icarus 143, 45-59.

\item{}
Richardson D.C., Michel P., Walsh, K.J. and K.W. Flynn 2009. Numerical simulations of asteroids modelled as gravitational aggregates 
with cohesion. Planetary and Space Science, in press.

\item{}
Stewart S.T., Leinhardt Z.M. 2009. Velocity-dependent catastrophic disruption criteria for planetesimals. 
Astrophys. J. Letters 691, L133-137.

\item{}
Tillotson, J.H., 1962. Metallic equations of state for hypervelocity
impact. General Atomic Report GA-3216, July 1962.

\item{}
Wilkison, S.L., Robinson, M.S., Thomas, P.C., Veverka, J., McCoy, T.J., Murchie, S.L., Prockter, L.M., Yeomans, D.K., 2002. 
An estimate of erosÕs porosity and implications for internal structure. Icarus 155, 94-103. 

\item{}
Yeomans, D.K., et al., 1997. Estimating the mass of asteroid 253 Mathilde from tracking data during the 
NEAR flyby. Science 278, 2106-2109.

\end{description}

\newpage



\clearpage

\clearpage

\begin{table}
\vskip 20pt
\vspace{2.truecm}
\begin{center}
\begin{tabular}{lll}
&Nominal Non-Porous&Nominal Porous \\
\hline
$m$&$9.5$&$9.5$ \\
$k$&$3.0 \times 10^{28}$&$8.0 \times 10^{37}$ \\
$Y$&$3.5 \times 10^{10}$&$3.5 \times 10^{10}$ \\
$\sigma_T(3 \rm cm)$&$3.2 \times 10^8$&$3.5 \times 10^7$ \\
\end{tabular}
\end{center}
\normalsize
\caption{Nominal material properties of non-porous and porous targets. $m$ and $k$ (in cm$^{-3}$) are the Weibull parameters 
used to characterize the distribution of incipient flaws. $Y$ (in dynes$/$cm$^2$) is the yield strength, and $\sigma_T(3 \rm cm)$ 
is the (size dependent) tensile strength (in dynes$/$cm$^2$) of a 3 cm-diameter target.
}
\label{t:nommatprop}
\end{table}

\begin{table}
\vskip 20pt
\vspace{2.truecm}
\begin{center}
\begin{tabular}{rccrr}
\multicolumn{1}{c}{$R_\textrm{t}$ (km)}&Type&
 \multicolumn{1}{c}{$R_\textrm{p}$ (km)} & \multicolumn{1}{c}{$Q$ (erg g$^{-1}$)} &
 $M_\mathrm{lr}/M_\mathrm{pb}$ \\
\hline
$0.00003$&NP&$0.0000022$ &$1.79 \times 10^7$&$0.49$ \\
$0.003$&NP&$0.00013$ &$3.34 \times 10^6$&$0.51$ \\
$0.3$&NP&$0.01$&$1.63 \times 10^6$&$0.54$ \\
$1.0$&NP&$0.058$&$8.38 \times 10^6$&$0.54$ \\
$10.0$&NP&$1.59$&$1.80 \times 10^8$&$0.46$ \\
$100.0$&NP&$41$&$3.10 \times 10^9$&$0.50$ \\
$0.00003$&P&$0.0000024$&$4.74 \times 10^7$&$0.50$ \\
$0.003$&P&$0.00011$&$4.49 \times 10^6$&$0.63$ \\
$0.3$&P&$0.0093$&$2.50 \times 10^6$&$0.51$ \\
$1.0$&P&$0.045$& $8.04  \times 10^6$ &$0.49$ \\
$10.0$&P&$1.01$&$9.55  \times 10^7$&$0.56$ \\
$100.0$&P&$27.3$&$1.90  \times 10^9$&$0.48$ \\
\end{tabular}
\end{center}
\normalsize
\caption{Summary of simulation parameters. NP and P refer to our nominal Non-Porous and 
Porous targets, respectively (see Table \ref{t:nommatprop}). The projectile's angle of
incidence is $45^\circ$ and the impact speed is $3$ km$/$s. Impact conditions 
are defined by the specific impact energy $Q$ = (projectile kinetic energy)/(target mass), which
involves the projectile's radius $R_\mathrm{p}$.
$M_\mathrm{lr}/M_\mathrm{pb}$ is the resulting mass ratio of the
 largest remnant to the parent body. All simulations are aimed at being
close to the catastrophic disruption threshold defined as
$M_\mathrm{lr}/M_\mathrm{pb}=0.5$. Note that for targets with a radius $R_t \ge 0.3$ km, the gravitational reaccumulation was explicitly simulated.
}
\label{t:nomic}
\end{table}

\begin{table}[ht!] \label{t:qcritfit}
\begin{center}
\vspace{0.5 truecm}
\begin{tabular}{lccccc}
Material&$v_{impact}$ (km/s)&$Q_0$ (erg/g)&$B$ (erg cm$^3$/g$^2$)&$a$&$b$\\
\hline
Porous (pumice)& 3 & 7.0 $\times$ 10$^7$ & 4.15 & -0.43& 1.22 \\
Porous (pumice)& 5 & 1.0 $\times$ 10$^8$& 5.70 & -0.45 & 1.22 \\
Non-porous (basalt)& 3 & 2.8 $\times$ 10$^7$ & 0.40 & -0.38 & 1.36 \\
Non-porous (basalt)& 5 & 2.9 $\times$ 10$^7$ & 1.50 & -0.35 & 1.29 \\
\end{tabular}
\caption{Fit constants for $Q_D^*$ (see text for details).}
\end{center}
\end{table}

\begin{table}
\vskip 20pt
\vspace{2.truecm}
\begin{center}
\begin{tabular}{lll}
&Weak Non-Porous&Strong Porous \\
\hline
$m$&$9.5$&$9.5$ \\
$k$&$8.0 \times 10^{37}$&$3.0 \times 10^{28}$ \\
$Y$&$1.0 \times 10^{8}$&$3.5 \times 10^{10}$ \\
$\sigma_T(3 \rm cm)$&$3.3 \times 10^7$&$3.5 \times 10^8$ \\
\end{tabular}
\end{center}
\normalsize
\caption{Same as Table \ref{t:nommatprop} for the weak non-porous and strong porous targets. Note that although the Weibull parameters of the weak non-porous targets 
are the same as the nominal ones of porous targets, the value of $\sigma_T(3 \rm cm)$ is slightly different; the reason is that the density of 
non-porous targets (and therefore the volume) is higher than the density of porous ones for a similar diameter.
}
\label{t:modmatprop}
\end{table}

\begin{table}\label{t:vej}
\vskip 20pt
\vspace{0.5truecm}
\begin{center}
\begin{tabular}{rccrrrr}
\multicolumn{1}{c}{$R_\textrm{T}$ (km)}&Type&
 \multicolumn{1}{c}{$R_\textrm{min}$ (km) } & \multicolumn{1}{c}{$V_\mathrm{lr}$
} &
 \multicolumn{1}{c}{$<\!V\!>$} & \multicolumn{1}{c}{$V_\mathrm{med}$}
 & \multicolumn{1}{c}{$V_\mathrm{max}$}
  \\ \hline
$0.3$&NP&$0.005$&$0.06$&$1.3$&$0.5$&$888$ \\
$1.0$&NP&$0.017$&$0.27$&$4.1$&$1.75$&$1606$ \\
$10.0$&NP&$0.172$&$5.5$&$34$&$17$&$3789$ \\
$100.0$&NP&$1.820$&$43$&$303$&$155$&$5562$ \\
$0.3$&P&$0.005$&$0.13$&$1.4$&$0.7$&$216$ \\
$1.0$&P&$0.016$&$0.4$&$7.3$&$3.9$&$695$ \\
$10.0$&P&$0.15$&$4.5$&$51$&$25$&$2759$ \\
$100.0$&P&$1.46$&$76$&$317$&$188$&$4876$ \\
\end{tabular}
\end{center}

\normalsize
\caption{Properties of fragment ejection speeds. NP and P refer to Non-Porous and Porous targets, respectively. 
$R_\textrm{min}$ is the radius of the smallest fragment in our simulations (all larger fragments underwent at least 
one reaccumulation event). Speeds are
given in m$/$s. $V_\mathrm{lr}$ is the largest remnant ejection
speed. $<\!V\!>$ is the average velocity of fragments which underwent
at least one reaccumulation event, while $V_\mathrm{med}$ and
$V_\mathrm{max}$ are, respectively, their median and maximum speed.
}
\end{table}

\end{document}